\documentclass[aps,prl,twocolumn,superscriptaddress,10pt]{revtex4-1}

\usepackage{amsmath}
\usepackage{amssymb}
\usepackage{graphicx}
\usepackage[utf8]{inputenc}
\usepackage{braket}
\usepackage[%
  colorlinks=true,
  urlcolor=blue,
  linkcolor=blue,
  citecolor=blue
]{hyperref}
\usepackage{color}
\usepackage[table]{xcolor}

\newcommand{\bs}[1]{\boldsymbol{#1}}

\newcommand{\fnr}{\ensuremath{\mathbf n_{\mathbf r}}}
\newcommand{\fr}{\ensuremath{\mathbf r}}
\newcommand{\fnrx}{\ensuremath{\mathbf n_{\mathbf r + \mathbf e_x}}}
\newcommand{\fnry}{\ensuremath{\mathbf n_{\mathbf r + \mathbf e_y}}}

\begin{document}

\title{Steering of the Skyrmion Hall Angle By Gate Voltages}

\author{J. Plettenberg}
\affiliation{I. Institut für Theoretische Physik, Universität Hamburg, Jungiusstraße 9, 20355 Hamburg, Germany}
\author{M. Stier}
\affiliation{I. Institut für Theoretische Physik, Universität Hamburg, Jungiusstraße 9, 20355 Hamburg, Germany}

\author{M. Thorwart}
\affiliation{I. Institut für Theoretische Physik, Universität Hamburg, Jungiusstraße 9, 20355 Hamburg, Germany}

\date{\today}

\begin{abstract}
Magnetic skyrmions can be driven by an applied spin-polarized electron current which exerts a spin-transfer torque on the localized spins constituting the skyrmion. However, the longitudinal dynamics is plagued by  
%
%
%
the skyrmion Hall effect which causes the skyrmions to acquire a transverse velocity component.
We show how to use spin-orbit interaction to control the skyrmion Hall angle and how 
the interplay of spin-transfer and spin-orbit torques can lead to a complete suppression of the transverse motion. 
Since the spin-orbit torques can be controlled all-electronically by a gate voltage, the skyrmion motion can be steered all-electronically on a broad racetrack at high speed and conceptually new writing and gating operations can be realized.

\end{abstract}


\maketitle 
Magnetic skyrmions (SKs) are topologically protected vortex-like spin textures that can be formed in non-centrosymmetric magnetic compounds. Due to their stability, their extremely small size, and the possibility to drive them by low current densities, they are promising candidates for spintronic devices such as racetrack memories. In crystals lacking spatial inversion symmetry, the interplay of Heisenberg exchange interaction, antisymmetric Dzyaloshinskii-Moriya interaction, and an external Zeeman field may lead to the formation of vortex-like magnetic SKs.  They have been predicted\cite{Bogdanov1989, Bogdanov1994, Bogdanov1995} years before they were experimentally discovered in magnetic layers with a strong spin-orbit interaction\cite{Muehlbauer2009, Yu2010, Heinze2011}.

SKs carry a nonzero, integer value topological charge $Q$, also called SK number\cite{Bergmann2014}. This number is an invariant that counts how many times the field configuration wraps around a unit sphere. It cannot be changed by continuous transformations. Due to this property, SKs are insusceptible to imperfect fabrication or disorder. On a lattice, the argument of topological stability has to be replaced by a finite energy barrier, but despite their small size which is typically about 10-100 nm\cite{Nagaosa2013}, SKs are quite stable. Due to the underlying emergent electromagnetic field induced by the Berry phase, SKs experience a Magnus force\cite{Manchon2014} that strongly suppresses pinning by deflecting SKs from pinning centers\cite{Lin2013}. Thus SKs can be driven at current densities of the order $10^5$ $\text{A}/\text{m}^2$, about four orders of magnitude lower than required, e.g., for domain walls\cite{Iwasaki2013, Lu2014}. This makes SKs very promising candidates for future spintronic applications, especially for racetrack memories consisting of thin nanowires.

For the use in technical applications, however, several hurdles have to be overcome. First, the creation of SKs needs to be possible. This has been demonstrated by various mechanisms, e.g., by sweeping external magnetic fields\cite{Koshibae2016} or by applying circular currents \cite{Tchoe2012}. 
In addition, the controlled creation and annihilation of single SKs has been realized \cite{Romming2013} and similar processes have been theoretically explained\cite{Stier2017,Everschor_Sitte_2017}. Direct  creation or annihilation  of SKs  suffers from the requirement of  large currents or fields. Current-driven SKs on a two-lane racetrack memory devices \cite{Mueller_2017,Suess_2018} have been proposed where SKs are placed on different ``lanes'' of a broad racetrack. However, current-driven SKs experience  the SK Hall effect, in which the SKs develop a motion perpendicular to the direction of the applied current, just like charged particles in the standard Hall effect\cite{Litzius2017, Jiang2017}. 
In experiments, the corresponding skyrmion Hall angle  $\Theta$ has exceeded 30$^\circ$\cite{Litzius2017}. 
It depends on the Gilbert damping, the nonadiabaticity parameter, and the spin torques, but is  independent of the external current density, at least when it overcomes some small threshold \cite{Jiang2017}. These facts have been previously explained based on a general SK equation of motion for the topological charge density \cite{Stier2017}, or the Thiele equation for a specific SK configuration \cite{Iwasaki2013}. A possible dependence on the external current density \cite{Litzius2017} is probably due to intrinsic pinning or SK deformation but is not yet fully explained.

The presence of the SK Hall effect limits the use of SKs on racetracks because the transverse velocity component can lead to annihilation of the SK at the edges of the track.  For this reason, the SK Hall effect is typically seen as a detrimental effect.  Several approaches have been proposed to keep SKs on the track. Most of them aim at creating potential barrier at the track edges, deflecting the SKs\cite{Kolesnikov2018}. However, this can lead to an inefficient and hard to control zigzag path and to irregular SK motion. 

In this work, we show that the Rashba spin-orbit interaction can be used to steer the SK Hall angle due to the interplay of  current-induced spin-transfer torques and Rashba spin-orbit torques. This can even be used to completely suppress the skyrmion Hall angle. Moreover, with an externally applied gate voltage it is possible to modify the magnitude of the spin-orbit torques allowing us to steer the skyrmion Hall angle all-electronically. With this mechanism it is possible to move skyrmions on a broad racetrack at high speed, to efficiently steer their trajectories, e.g., to change lanes,   and to realize conceptually new writing and gating operations with a tunable gate voltage.

{\it Model -- }From a theoretical point of view, SKs are two-dimensional quasiparticles that obey the Landau-Lifshitz-Gilbert equation \cite{Landau1935, Landau1960, Gilbert2004, Nakatani1989}, a partial differential equation describing the precessional motion of magnetic moments  in a ferromagnetic material. To describe current driven SKs, it is extended by adiabatic and nonadiabatic spin torques, $\bs T^\text{ad}$ and $\bs T^\text{nonad}$, which are induced by spin-polarized currents\cite{Slonczewski1996, Bazaliy1998, Zhang2004}, and reads
\begin{equation} \label{LLG}
\partial_t \bs n = - \bs n \times \bs B_\text{eff} + \alpha \bs n \times \partial_t \bs n
+ \bs T^\text{ad} + \bs T^\text{nonad}
\end{equation}
with the normalized magnetization vector field $\bs n =  \bs n(x,y,t)$\cite{Nagaosa2013} and the Gilbert damping constant $\alpha>0$. The effective field $\bs B_\text{eff} = - \partial H / \partial \bs n$ contains all interactions of the system Hamiltonian $H$.
Here, the gyromagnetic ratio $\gamma$ is absorbed in $\bs B_\text{eff}$, $\bs T^\text{ad}$, and $\bs T^\text{nonad}$ and we set $\hbar=1$. 

{\em Current-induced spin torques -- } We have calculated the current-induced spin-torques up to second order in the Rashba spin-orbit coupling parameter $\alpha_\text{R}$ based on a semi-classical Boltzmann approach \cite{SM}.
To zeroth order in $\alpha_\text{R}$, the adiabatic spin-transfer torque 
\begin{equation}
\bs T^\text{ad}_\text{STT} = v_\text{s} \partial_x \bs n
\end{equation}
is recovered. The prefactor $v_\text{s} = P a^3 j_e / (2 e)$ with spin polarization $P$, lattice constant $a$ and elementary charge $e$ has the dimension of a velocity and is called effective spin velocity. The effective spin velocity is proportional to the external current density $j_e$ and can therefore easily be tuned.
In addition to the spin-transfer torque, we find the adiabatic first-order spin-orbit torque
\begin{equation}
\bs T^\text{ad}_\text{SOT} =
\underbrace{\frac{2 m \alpha_\text{R}}{\hbar^2} }_{\equiv\lambda^{-1}}v_\text{s} (\bs n \times \hat{\bs y})\label{eq::SOT}
\end{equation}
with the effective electron mass $m$ and the inverse spin-orbit length $\lambda^{-1}$.
Up to first order, all relevant torques reported in the literature \cite{Zhang2004, Manchon2009,vanderBijl2012, Stier2013, Stier2014, Stier2015} are recovered. 

For the sake of simplicity, we neglect second order spin-orbit torques in the following discussion, thus $\bs T^\text{ad} \approx \bs T^\text{ad}_\text{STT} + \bs T^\text{ad}_\text{SOT}$. As demonstrated in the Supplemental Material \cite{SM}, second order torques can enhance the effects discussed in this work and should be considered under certain circumstances.

Damping of the spin dynamics of the localized electrons is described by the Gilbert damping term.
Due to effects like impurity scattering or spin-orbit coupling, the itinerant electrons experience damping as well. The corresponding nonadiabatic spin torques are obtained as 
$\bs T^\text{nonad}= - \beta \bs n \times \bs T^\text{ad}$ \cite{vanderBijl2012} with the nonadiabaticity parameter $\beta$. Since spin-orbit coupling is one of the main damping sources, the nonadiabatic spin-orbit torques can play a major role for the skyrmion dynamics. 
This property is ultimately responsible for the possibility to steer SKs.

{\em Controlling the Skyrmion Hall angle -- }
The topological charge of a SK leads to a theoretically predicted SK Hall effect, in which the SKs acquire a velocity component perpendicular to the applied electronic current.  The SK dynamics is governed by Eq.\ (\ref{LLG}) and is illustrated for the SK Hamiltonian
\begin{align}
\label{eq:model_Hamiltonian}
H = & -J \sum_{\bs r} \bs n_{\bs r} \cdot
[\bs n_{\bs r + a \hat{\bs x}} + \bs n_{\bs r + a \hat{\bs y}}]
- \bs B \sum_{\bs r} \bs n_{\bs r}
\nonumber\\
& -D \sum_{\fr}\Big[\hat{\mathbf y}\cdot(\fnr\times\fnrx)+\hat{\mathbf x}\cdot(\fnr\times\fnry)\Big]
\end{align} 
defined on a square lattice with lattice sites ${\bs r}=(x,y)$ and a fixed lattice constant $a=0.5$ nm.
$J=1$ meV is the exchange interaction, $\bs B = - 0.03 \text{ eV } \hat{\bs z}$ (corresponding to $\sim - 0.5$ T) an external Zeeman field, and $D = 0.36$ meV the Dzyaloshinskii-Moriya interaction strength. The choice of an interfacial Dzyaloshinskii-Moriya interaction stabilizes N\'eel SKs with a radius of approximately 5nm which is approximately the observed size in SrIrO$_3$/SrRuO$_3$ layers.
Typical experimental values of the Gilbert damping parameter $\alpha$ and nonadiabaticity parameter $\beta$ cover a wide range. For $\alpha$, values ranging from $\alpha < 0.01$ to $\alpha \approx 1$\cite{Thomas2006, Koetzler2007} and for $\beta$, values ranging from $\beta \approx 0.02$ up to $\beta > 4$\cite{Sekiguchi2012, Martinez2008} have been reported for various materials.

As shown in the following, the SK Hall angle $\Theta$ can be steered by the use of the spin-orbit torque which is generated by the spin-orbit interaction in the itinerant electrons.  This clearly depends on the magnitude of the spin-orbit torque which is proportional to the Rashba coupling constant $\alpha_R$ and thus $\lambda^{-1}$. Since $\lambda^{-1}$ can be tuned relatively easy by gate voltages, we obtain the possibility to eventually steer $\Theta$ all-electronically. For many materials with either bulk or interfacial inversion symmetry breaking, a wide range of values  for the phenomenological Rashba parameter $\alpha_\text{R}$ has been reported. Furthermore, it has been demonstrated experimentally that by applying an external gate voltage, the structure inversion symmetry of a crystalline lattice can be lifted, leading to variations of $\alpha_\text{R}$ in the range of $10^{-12}$ to $10^{-11}$ eVm \cite{Ho2013,Caviglia2010_tunableRashba} which equates to $\lambda^{-1} \approx (0.01\dots 0.15)/a$ ($a$ lattice constant) when we use the free electron mass in Eq. (\ref{eq::SOT}). Notably, this also works for metal interfaces as recent reports show \cite{chen2018electric,soimodification2014}. The most promising class of materials should be, however, quasi 2D systems including LaAlO$_3$/SrTiO$_3$ (LAO/STO), SrIrO$_3$/SrRuO$_3$ (SIO/STO) or SrRuO$_3$/SrTiO$_3$ (SRO/STO). In LAlO/STO can be efficiently tuned by a gate voltage \cite{yin2019tuning,narayanapillai2014current,lin2019interface,lesne2016highly}. In SRO/STO there is a rather large spin-orbit coupling as well as a damping of $\alpha\approx0.7\dots0.9$ \cite{langner2009ultrafast,RevModPhys.84.253}. In SIO/STO nanoscale SKs have been detected with a radius of about 6nm \cite{nanolett19}. In addition, the presence of SKs appears to be gate-controlled \cite{ohuchi2018electric} also hinting that the spin-orbit coupling is gate-controlled as well \cite{gu2018oxygen,ohuchi2018electric}.\\
Thus,  suitable materials containing SKs and a gate-controllable, sufficient Rashba coupling seem within reach. As the interest in these materials regarding SKs is quite new, not all parameters are determined by experiments and we cannot exclude an influence of the gate voltage on additional interactions as the DMI\cite{gu2018oxygen}. But we expect the parameters to stay in the range of the literature given above, which we adopt for our numerical simulations.

\begin{figure}[tb]
\centering
\includegraphics[width=0.99\linewidth]{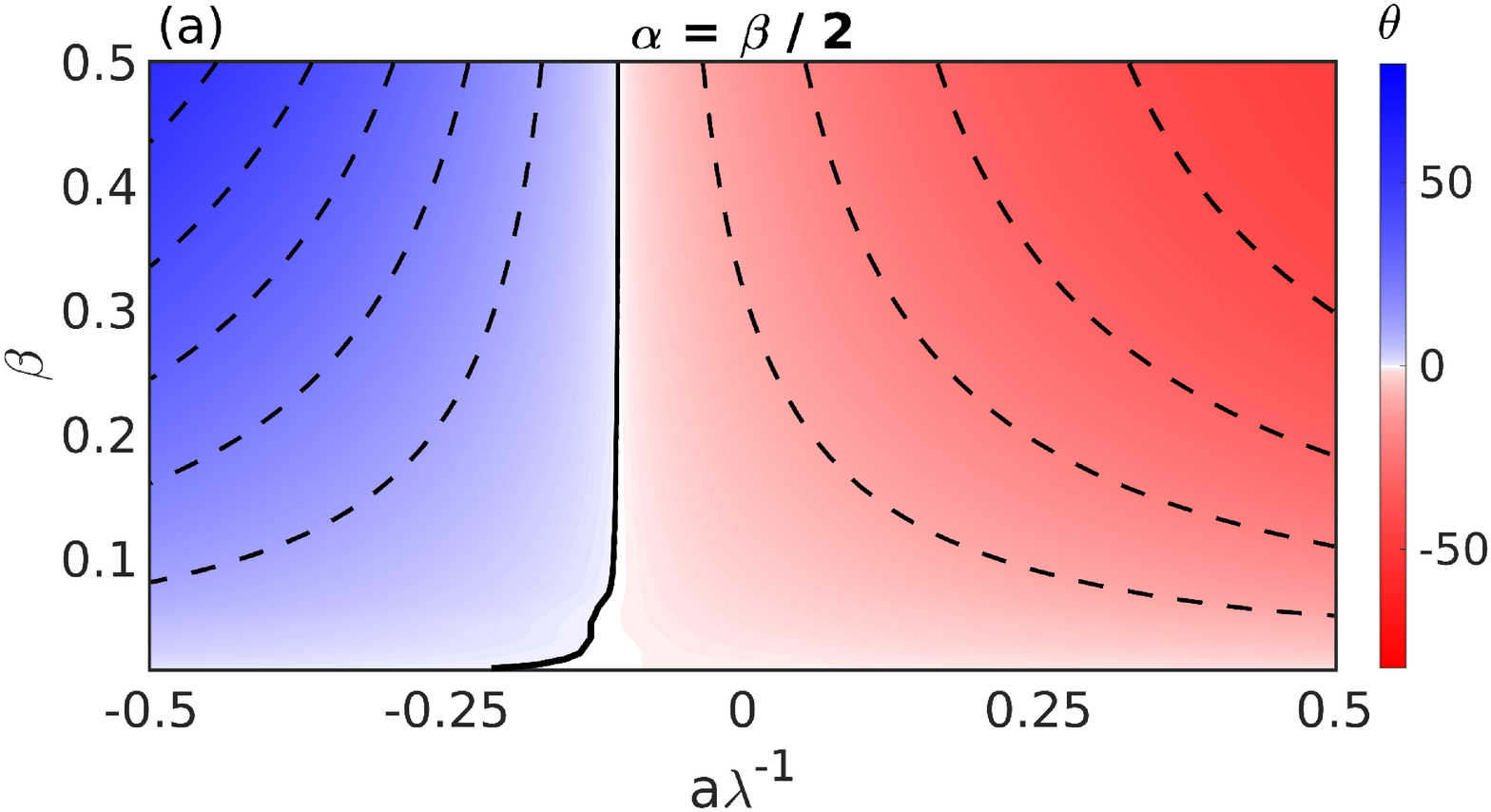}\\
\includegraphics[width=0.99\linewidth]{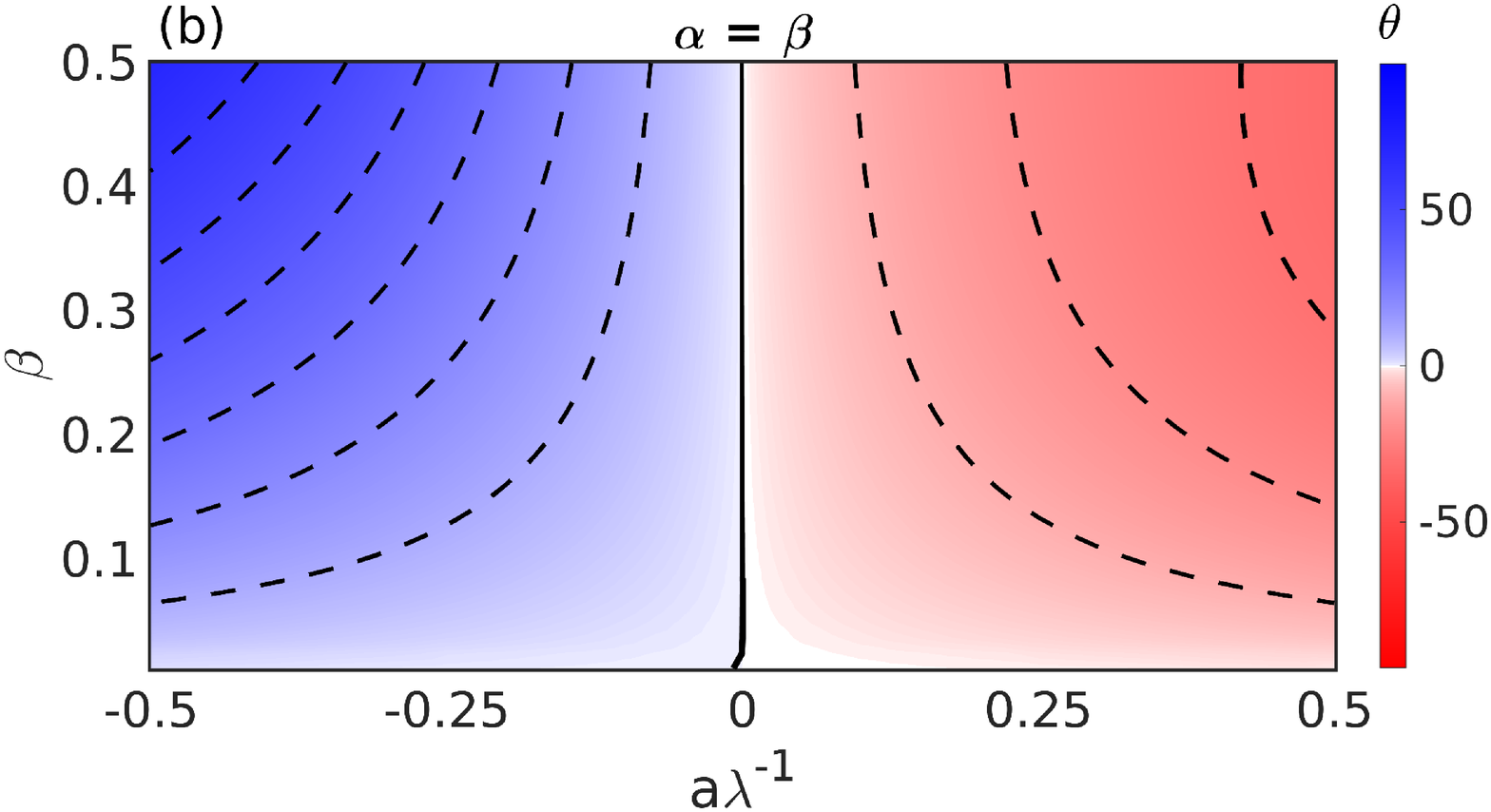}\\
\includegraphics[width=0.99\linewidth]{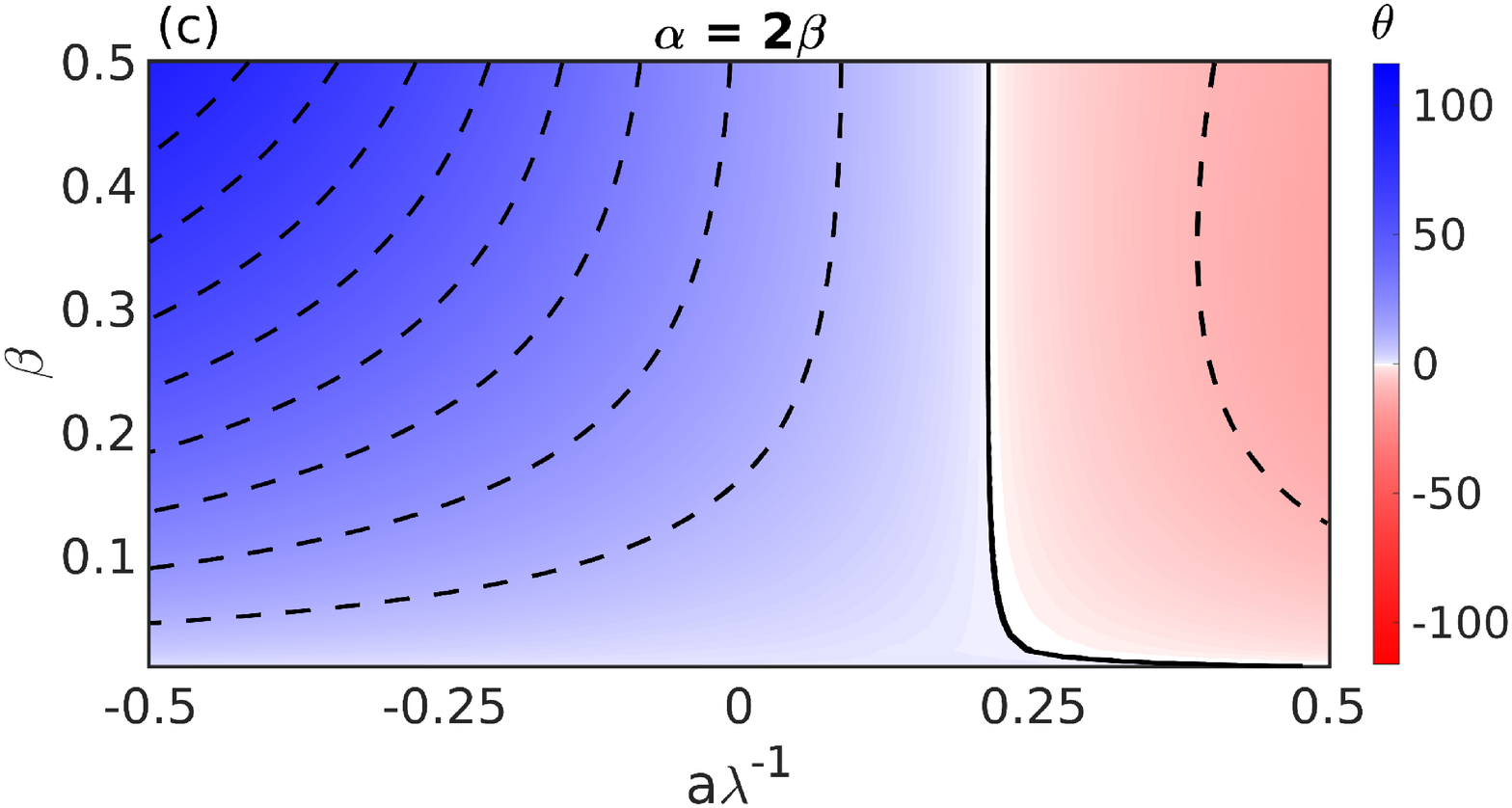}
\caption{\label{fig::SkH}
Skyrmion Hall angle  $\Theta$ as a function of the Rashba spin-orbit length $\lambda$ (in units of the lattice constant)  and $\beta$ while keeping the ratio $\beta/\alpha$ constant at (a) $\alpha=\beta/2$, (b) $\alpha=\beta$, and (c) $\alpha=2\beta$. Blue colors confer to positive and red colors to negative values. We included dashed contours every $10^{\circ}$ as a guide for the eyes while the solid line marks the vanishing of the SK Hall angle at $\Theta=0^{\circ}$. $j_e=10^9$A/m$^2$
}
\end{figure}
To illustrate the idea of a Rashba controlled SK Hall angle by explicit results, we show the effect of $\lambda^{-1}=2 m \alpha_\text{R}/\hbar^2$ and $\beta$ on $\Theta$ for three parameter choices of $\alpha / \beta$ in Fig. \ref{fig::SkH}. As for a vanishing Rashba interaction $\lambda^{-1} = 0$ the Skyrmion Hall angle $\Theta\propto (\alpha -\beta)$ \cite{Stier2017} we find $\Theta(\lambda^{-1}=0) < 0$ if $\alpha > \beta$, $\Theta(\lambda^{-1}=0) > 0$ if $\alpha < \beta$, and $\Theta(\lambda^{-1}=0) = 0$ if $\alpha = \beta$. When we add a finite Rashba interaction to the system we basically introduce a spin torque which acts as an in-plane magnetic field and breaks the symmetry. The SK Hall angle will thus be changed depending on the geometry of the Skyrmion and the explicit direction of Rashba induced SOT. For the system treated in this work, i.e. N\'eel SK, Rashba field in $\hat {\mathbf y}$ direction, current flow in $x$ direction, an increase of $\lambda^{-1}$ will decrease the SK Hall angle. Thus, for the simplest case $\alpha=\beta$ positive values of $\lambda^{-1}$ will yield a negative $\Theta$ and vice versa. Remarkably, due to the symmetry breaking created by the choice of a certain direction of the current flow and the direction of the SOT, the values of the SK Hall angle are not symmetric w.r.t. $\lambda^{-1} =0$.

\begin{figure}[tb]
\includegraphics[width=0.99\linewidth]{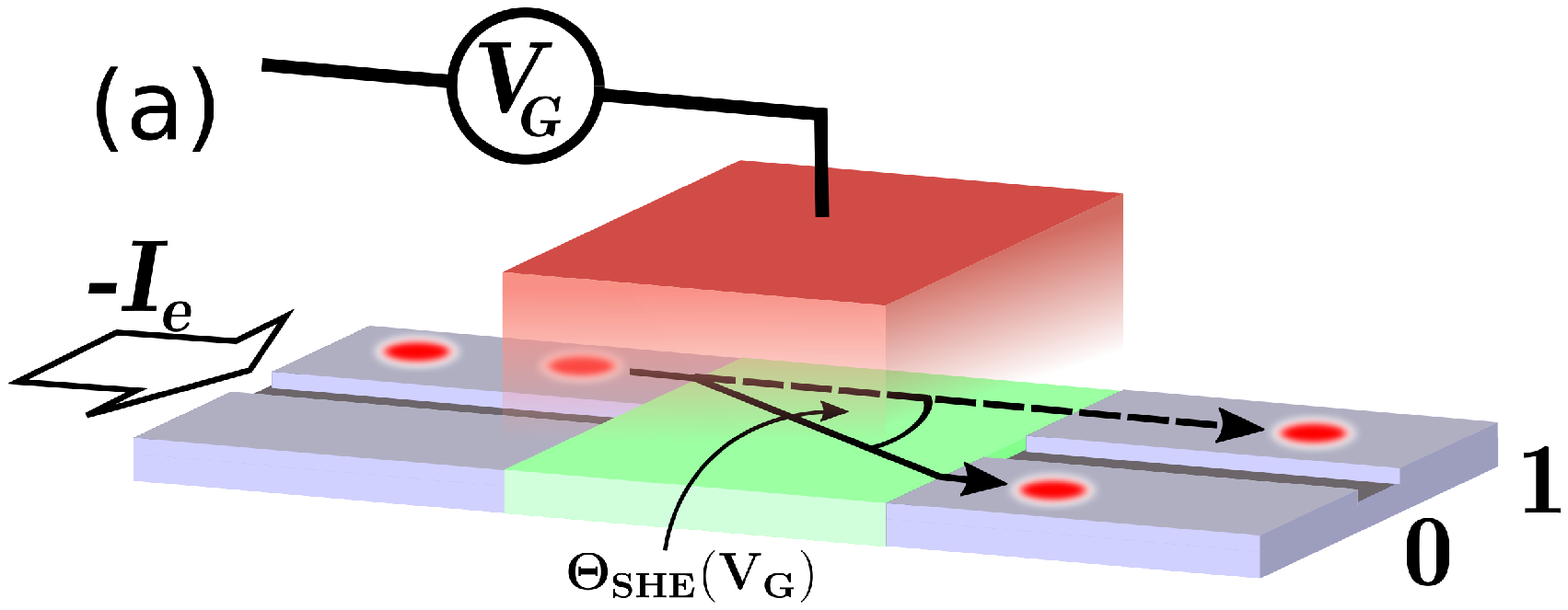}\\
\includegraphics[width=0.99\linewidth]{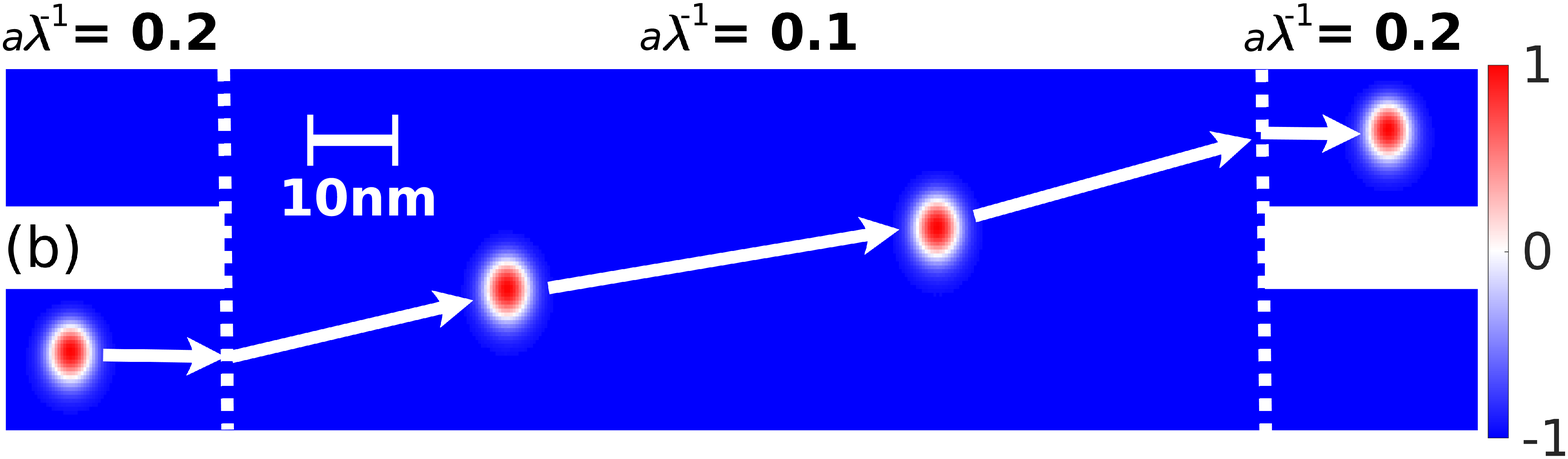}\\ 
\includegraphics[width=0.99\linewidth]{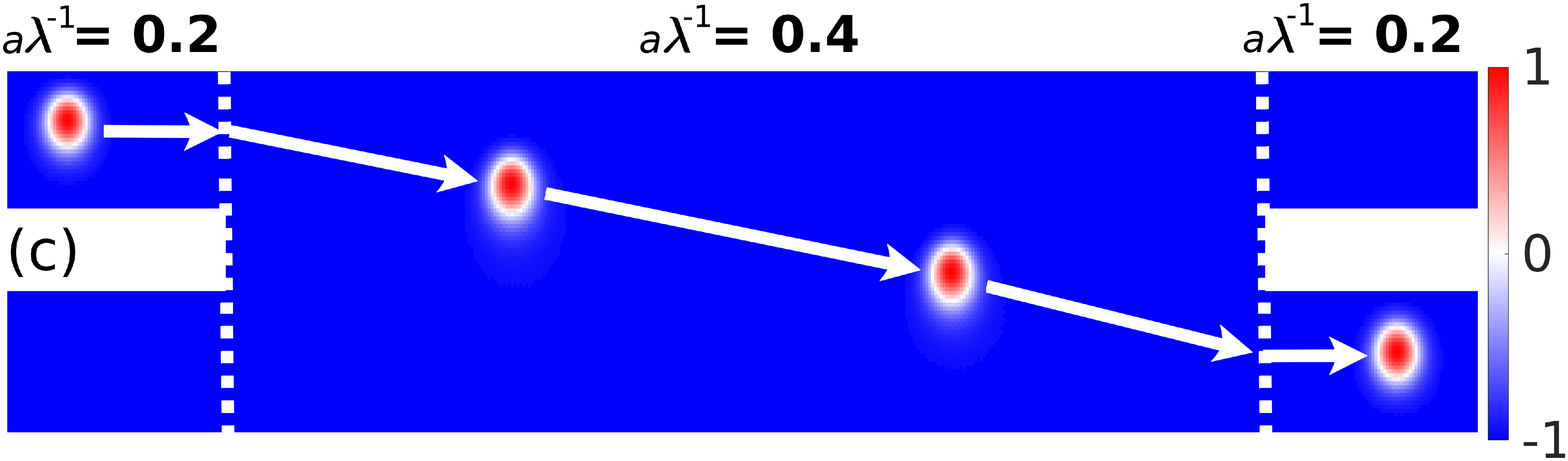}\\ 
\includegraphics[width=0.99\linewidth]{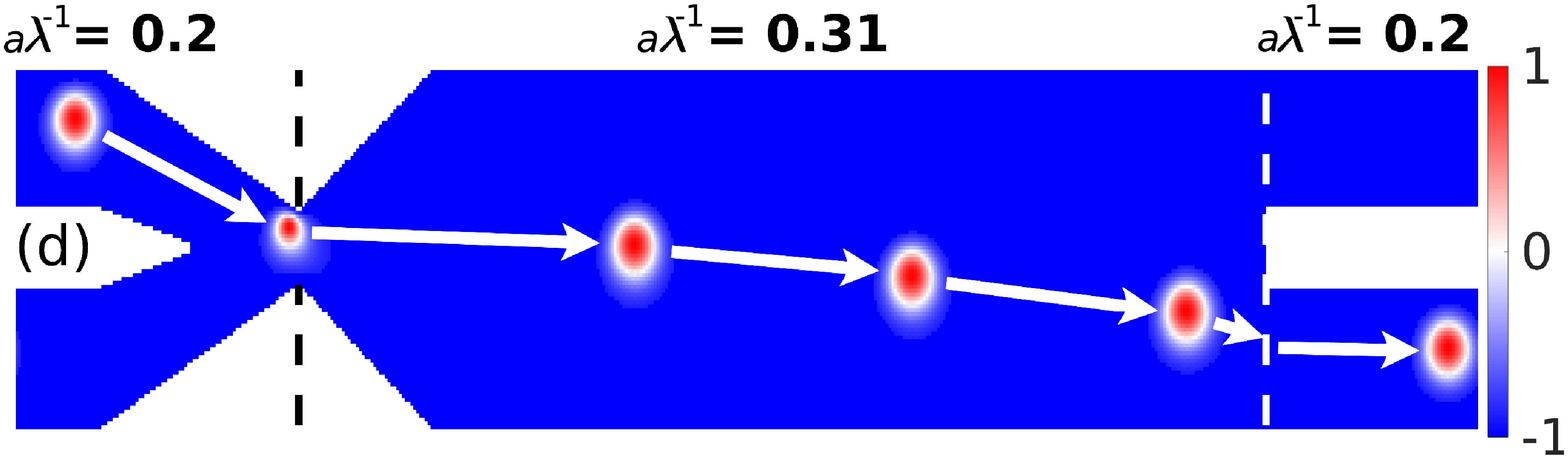}
\caption{\label{fig::broad_racetrack} 
(a) Sketch of a all-electronically controlled two-lane SK racetrack. Incoming SKs (red circles) get driven by an external spin polarized current $I_e$. In the blue regions, SKs are located either in the left or in the right lane of the track, the former corresponding to a logical "1" and the latter to a "0".
In the writing area (green), $\lambda^{-1}$ is altered by an external gate voltage $V_G$ leading to a tunable $\Theta$. In this area, the two lanes are not separated anymore. Therefore incoming SKs can be transferred between the two lanes. After leaving the writing area, the initial configuration is restored and SKs move along their lanes.\\
(b), (c), and (d): Magnetization $n_z$ from micromagnetic simulations for the indicated values of $\lambda^{-1}$ (in units of inverse lattice constant) in the respective areas separated by the dashed lines. Snapshots of a single SK's trajectory are taken every (b) $\Delta t = 5$ns, (c) $\Delta t = 4.5$ns, or $\Delta t = 3$ns. Skyrmions will move straight only within the lanes but at an angle in the switching area as indicated by the white arrows. The advanced geometry in (d) allows switching for a smaller Rashba coupling.  $\alpha=2\beta=0.4$, $j_e =1.5\times 10^{10}$A/m$^2$, sample size $48\text{nm}\times 256$nm
}
\end{figure}

{\em Concept of a Skyrmion racetrack gate --}
The possibility of an all-electronic steering of the SK Hall angle opens the doorway to the conceptual design of a gate in a  SK two-lane racetrack, as it is sketched in Fig. \ref{fig::broad_racetrack}a. 
%
%
A SK can either be located in the left or in the right lane of the racetrack, the former corresponding to a "1" and the latter to a "0". 
The racetrack could work as follows. A spin polarized current is applied to a broad quantum wire. By choosing suitable materials and possibly applying a gate voltage, the magnitude of the Rashba parameter is adapted such that for a running SK,  $\Theta=0^{\circ}$  and the SK can move along one lane of the wire on straight trajectories (blue areas in Fig. \ref{fig::broad_racetrack}a) and keep them in their lanes.  
Only within the writing area the two lanes are connected.
In this writing area, $\alpha_\text{R}$ gets altered by a different gate voltage leading to a tunable Skyrmion Hall angle. Thus, the SK can change between the two lanes. After leaving the writing area, the initial configuration is restored and the SK moves again in the direction of the current along the track.
With this mechanism, SK movement on the track at high speed is possible as well as a logical switching of SKs without the need to annihilate or create them. Reading operations can be realized using the well-known effect of the giant magneto-resistance.
As a proof of concept we also show the results of micromagnetic simulations in Fig. \ref{fig::broad_racetrack}. We adapt the basic idea of Fig. \ref{fig::broad_racetrack}a by setting different values of $\lambda^{-1}$ in the lanes and the switching area. Due to the different cross sections of the conductor in the respective areas, the current flow has to be calculated by a solution of the Laplace equation $\Delta \phi = 0$ where $\Delta$ is the Laplace operator and $\phi$ the electrical potential. We apply von-Neumann boundary conditions to ensure a current flow of $v_s$ at the left and right edges of the sample while the remaining edges are insulating. Choosing $\alpha = 2\beta=0.4$ we adopt the situation of Fig. \ref{fig::SkH}c which allows for a switching of the sign of the SK Hall angle around values of $\lambda^{-1} \approx 0.2/a$. Subfigures \ref{fig::broad_racetrack}b and \ref{fig::broad_racetrack}c in fact show a switching of the lanes when we apply $\lambda^{-1} = 0.1/a$ or $\lambda^{-1} = 0.4/a$, respectively. The choice of $\lambda^{-1} =0.2/a$ within the lanes lets the SKs move straight and thus most effectively. As we have seen from Fig. \ref{fig::SkH}c the dependence of $\Theta$ on $\lambda^{-1}$ is non-linear and typically harder to switch in a certain direction. For the given parameters this would be downwards as in Fig. \ref{fig::broad_racetrack}c. The necessary Rashba coupling can, however, be reduced by an advanced geometry as shown in Fig. \ref{fig::broad_racetrack}d. Here, a constriction already pushes the SK downwards and thus reduces the necessary Rashba coupling to $\lambda^{-1} = 0.31/a$. As modern manufacturing methods allow for a controlled creation of far more complex geometries, this opens the concept of spin-orbit induced switching of SKs to a wider class of materials where the range of the gate-controlled Rashba coupling $\alpha_R$ would be too small for simple geometries.


{\em Conclusions --} In this work, we have proposed a mechanism to steer the dynamics of current-driven SKs by all-electronically controlling the SK Hall angle by an external gate voltage.
As a basis for this, we have derived the expressions for spin-transfer and spin-orbit torques based on the semi-classical Boltzmann equation in two spatial dimensions. We show that the SK Hall angle is then controlled by the interplay of the spin-transfer torque and the nonadiabatic spin-orbit torque where the magnitude of the spin-orbit torque is proportional to the Rashba spin-orbit coupling parameter. As this parameter can be easily tuned by external gate voltages, it is possible to manipulate the ratio of spin-transfer and spin-orbit torques by applying a suitable gate voltage. This provides a simple and controlled way to steer the SKs dynamics. Based on this finding, we have sketched the design of a broad SK two-lane racetrack on which SKs can be moved at high speed in the desired direction. In addition, by tuning the Rashba spin-orbit coupling in a spatially restricted region, we have proposed the design of a SK gate for the two-lane racetrack. Due to the incapability of previous approaches to control the SK Hall effect, various technological applications based on SKs have been hard to realize and SK based racetracks have not yet been experimentally realized. We have verified our theoretical predictions by numerical simulations of SKs stabilized by the Dzyaloshinskii-Moriya interaction in the bulk on a square lattice.

{\em Acknowledgements --} J. Plettenberg and M. Stier contributed equally to this work. We gratefully acknowledge funding by the Deutsche Forschungsgemeinschaft (project number 403505707) within the SPP 2137 ''Skyrmionics''.





\begin{thebibliography}{54}%
\makeatletter
\providecommand \@ifxundefined [1]{%
 \@ifx{#1\undefined}
}%
\providecommand \@ifnum [1]{%
 \ifnum #1\expandafter \@firstoftwo
 \else \expandafter \@secondoftwo
 \fi
}%
\providecommand \@ifx [1]{%
 \ifx #1\expandafter \@firstoftwo
 \else \expandafter \@secondoftwo
 \fi
}%
\providecommand \natexlab [1]{#1}%
\providecommand \enquote  [1]{``#1''}%
\providecommand \bibnamefont  [1]{#1}%
\providecommand \bibfnamefont [1]{#1}%
\providecommand \citenamefont [1]{#1}%
\providecommand \href@noop [0]{\@secondoftwo}%
\providecommand \href [0]{\begingroup \@sanitize@url \@href}%
\providecommand \@href[1]{\@@startlink{#1}\@@href}%
\providecommand \@@href[1]{\endgroup#1\@@endlink}%
\providecommand \@sanitize@url [0]{\catcode `\\12\catcode `\$12\catcode
  `\&12\catcode `\#12\catcode `\^12\catcode `\_12\catcode `\%12\relax}%
\providecommand \@@startlink[1]{}%
\providecommand \@@endlink[0]{}%
\providecommand \url  [0]{\begingroup\@sanitize@url \@url }%
\providecommand \@url [1]{\endgroup\@href {#1}{\urlprefix }}%
\providecommand \urlprefix  [0]{URL }%
\providecommand \Eprint [0]{\href }%
\providecommand \doibase [0]{http://dx.doi.org/}%
\providecommand \selectlanguage [0]{\@gobble}%
\providecommand \bibinfo  [0]{\@secondoftwo}%
\providecommand \bibfield  [0]{\@secondoftwo}%
\providecommand \translation [1]{[#1]}%
\providecommand \BibitemOpen [0]{}%
\providecommand \bibitemStop [0]{}%
\providecommand \bibitemNoStop [0]{.\EOS\space}%
\providecommand \EOS [0]{\spacefactor3000\relax}%
\providecommand \BibitemShut  [1]{\csname bibitem#1\endcsname}%
\let\auto@bib@innerbib\@empty
\bibitem [{\citenamefont {Bogdanov}\ and\ \citenamefont
  {Yablonskii}(1989)}]{Bogdanov1989}%
  \BibitemOpen
  \bibfield  {author} {\bibinfo {author} {\bibfnamefont {A.~N.}\ \bibnamefont
  {Bogdanov}}\ and\ \bibinfo {author} {\bibfnamefont {D.~A.}\ \bibnamefont
  {Yablonskii}},\ }\href
  {http://www.jetp.ac.ru/cgi-bin/e/index/e/68/1/p101?a=list} {\bibfield
  {journal} {\bibinfo  {journal} {Zh. Eksp. Teor. Fiz}\ }\textbf {\bibinfo
  {volume} {95}},\ \bibinfo {pages} {178} (\bibinfo {year} {1989})}\BibitemShut
  {NoStop}%
\bibitem [{\citenamefont {Bogdanov}\ and\ \citenamefont
  {Hubert}(1994)}]{Bogdanov1994}%
  \BibitemOpen
  \bibfield  {author} {\bibinfo {author} {\bibfnamefont {A.}~\bibnamefont
  {Bogdanov}}\ and\ \bibinfo {author} {\bibfnamefont {A.}~\bibnamefont
  {Hubert}},\ }\href {\doibase 10.1016/0304-8853(94)90046-9} {\bibfield
  {journal} {\bibinfo  {journal} {J. Magn. Magn. Mater.}\ }\textbf {\bibinfo
  {volume} {138}},\ \bibinfo {pages} {255} (\bibinfo {year}
  {1994})}\BibitemShut {NoStop}%
\bibitem [{\citenamefont {Bogdanov}(1995)}]{Bogdanov1995}%
  \BibitemOpen
  \bibfield  {author} {\bibinfo {author} {\bibfnamefont {A.~N.}\ \bibnamefont
  {Bogdanov}},\ }\href {http://jetpletters.ac.ru/ps/1214/article_18359.shtml}
  {\bibfield  {journal} {\bibinfo  {journal} {JETP Lett.}\ }\textbf {\bibinfo
  {volume} {62}},\ \bibinfo {pages} {247} (\bibinfo {year} {1995})}\BibitemShut
  {NoStop}%
\bibitem [{\citenamefont {Mühlbauer}\ \emph {et~al.}(2009)\citenamefont
  {Mühlbauer}, \citenamefont {Binz}, \citenamefont {Jonietz}, \citenamefont
  {Pfleiderer}, \citenamefont {Rosch}, \citenamefont {Neubauer}, \citenamefont
  {Georgii},\ and\ \citenamefont {Böni}}]{Muehlbauer2009}%
  \BibitemOpen
  \bibfield  {author} {\bibinfo {author} {\bibfnamefont {S.}~\bibnamefont
  {Mühlbauer}}, \bibinfo {author} {\bibfnamefont {B.}~\bibnamefont {Binz}},
  \bibinfo {author} {\bibfnamefont {F.}~\bibnamefont {Jonietz}}, \bibinfo
  {author} {\bibfnamefont {C.}~\bibnamefont {Pfleiderer}}, \bibinfo {author}
  {\bibfnamefont {A.}~\bibnamefont {Rosch}}, \bibinfo {author} {\bibfnamefont
  {A.}~\bibnamefont {Neubauer}}, \bibinfo {author} {\bibfnamefont
  {R.}~\bibnamefont {Georgii}}, \ and\ \bibinfo {author} {\bibfnamefont
  {P.}~\bibnamefont {Böni}},\ }\href {\doibase 10.1126/science.1166767}
  {\bibfield  {journal} {\bibinfo  {journal} {Science}\ }\textbf {\bibinfo
  {volume} {323}},\ \bibinfo {pages} {915} (\bibinfo {year}
  {2009})}\BibitemShut {NoStop}%
\bibitem [{\citenamefont {Yu}\ \emph {et~al.}(2010)\citenamefont {Yu},
  \citenamefont {Onose}, \citenamefont {Kanazawa}, \citenamefont {Park},
  \citenamefont {Han}, \citenamefont {Matsui}, \citenamefont {Nagaosa},\ and\
  \citenamefont {Tokura}}]{Yu2010}%
  \BibitemOpen
  \bibfield  {author} {\bibinfo {author} {\bibfnamefont {X.~Z.}\ \bibnamefont
  {Yu}}, \bibinfo {author} {\bibfnamefont {Y.}~\bibnamefont {Onose}}, \bibinfo
  {author} {\bibfnamefont {N.}~\bibnamefont {Kanazawa}}, \bibinfo {author}
  {\bibfnamefont {J.~H.}\ \bibnamefont {Park}}, \bibinfo {author}
  {\bibfnamefont {J.~H.}\ \bibnamefont {Han}}, \bibinfo {author} {\bibfnamefont
  {Y.}~\bibnamefont {Matsui}}, \bibinfo {author} {\bibfnamefont
  {N.}~\bibnamefont {Nagaosa}}, \ and\ \bibinfo {author} {\bibfnamefont
  {Y.}~\bibnamefont {Tokura}},\ }\href {\doibase 10.1038/nature09124}
  {\bibfield  {journal} {\bibinfo  {journal} {Nature}\ }\textbf {\bibinfo
  {volume} {465}},\ \bibinfo {pages} {901} (\bibinfo {year}
  {2010})}\BibitemShut {NoStop}%
\bibitem [{\citenamefont {Heinze}\ \emph {et~al.}(2011)\citenamefont {Heinze},
  \citenamefont {von Bergmann}, \citenamefont {Menzel}, \citenamefont {Brede},
  \citenamefont {Kubetzka}, \citenamefont {Wiesendanger}, \citenamefont
  {Bihlmayer},\ and\ \citenamefont {Blügel}}]{Heinze2011}%
  \BibitemOpen
  \bibfield  {author} {\bibinfo {author} {\bibfnamefont {S.}~\bibnamefont
  {Heinze}}, \bibinfo {author} {\bibfnamefont {K.}~\bibnamefont {von
  Bergmann}}, \bibinfo {author} {\bibfnamefont {M.}~\bibnamefont {Menzel}},
  \bibinfo {author} {\bibfnamefont {J.}~\bibnamefont {Brede}}, \bibinfo
  {author} {\bibfnamefont {A.}~\bibnamefont {Kubetzka}}, \bibinfo {author}
  {\bibfnamefont {R.}~\bibnamefont {Wiesendanger}}, \bibinfo {author}
  {\bibfnamefont {G.}~\bibnamefont {Bihlmayer}}, \ and\ \bibinfo {author}
  {\bibfnamefont {S.}~\bibnamefont {Blügel}},\ }\href {\doibase
  10.1038/NPHYS2045} {\bibfield  {journal} {\bibinfo  {journal} {Nat. Phys.}\
  }\textbf {\bibinfo {volume} {7}},\ \bibinfo {pages} {713} (\bibinfo {year}
  {2011})}\BibitemShut {NoStop}%
\bibitem [{\citenamefont {von Bergmann}\ \emph {et~al.}(2014)\citenamefont {von
  Bergmann}, \citenamefont {Kubetzka}, \citenamefont {Pietzsch},\ and\
  \citenamefont {Wiesendanger}}]{Bergmann2014}%
  \BibitemOpen
  \bibfield  {author} {\bibinfo {author} {\bibfnamefont {K.}~\bibnamefont {von
  Bergmann}}, \bibinfo {author} {\bibfnamefont {A.}~\bibnamefont {Kubetzka}},
  \bibinfo {author} {\bibfnamefont {O.}~\bibnamefont {Pietzsch}}, \ and\
  \bibinfo {author} {\bibfnamefont {R.}~\bibnamefont {Wiesendanger}},\ }\href
  {\doibase 10.1088/0953-8984/26/39/394002} {\bibfield  {journal} {\bibinfo
  {journal} {J. Phys.: Condens. Matter}\ }\textbf {\bibinfo {volume} {26}},\
  \bibinfo {pages} {394002} (\bibinfo {year} {2014})}\BibitemShut {NoStop}%
\bibitem [{\citenamefont {Nagaosa}\ and\ \citenamefont
  {Tokura}(2013)}]{Nagaosa2013}%
  \BibitemOpen
  \bibfield  {author} {\bibinfo {author} {\bibfnamefont {N.}~\bibnamefont
  {Nagaosa}}\ and\ \bibinfo {author} {\bibfnamefont {Y.}~\bibnamefont
  {Tokura}},\ }\href {\doibase 10.1038/nnano.2013.243} {\bibfield  {journal}
  {\bibinfo  {journal} {Nat Nano}\ }\textbf {\bibinfo {volume} {8}},\ \bibinfo
  {pages} {899} (\bibinfo {year} {2013})}\BibitemShut {NoStop}%
\bibitem [{\citenamefont {Manchon}(2014)}]{Manchon2014}%
  \BibitemOpen
  \bibfield  {author} {\bibinfo {author} {\bibfnamefont {A.}~\bibnamefont
  {Manchon}},\ }\href {\doibase 10.1038/nphys2957} {\bibfield  {journal}
  {\bibinfo  {journal} {Nat. Phys.}\ }\textbf {\bibinfo {volume} {10}},\
  \bibinfo {pages} {340} (\bibinfo {year} {2014})}\BibitemShut {NoStop}%
\bibitem [{\citenamefont {Lin}\ \emph {et~al.}(2013)\citenamefont {Lin},
  \citenamefont {Reichhardt}, \citenamefont {Batista},\ and\ \citenamefont
  {Saxena}}]{Lin2013}%
  \BibitemOpen
  \bibfield  {author} {\bibinfo {author} {\bibfnamefont {S.-Z.}\ \bibnamefont
  {Lin}}, \bibinfo {author} {\bibfnamefont {C.}~\bibnamefont {Reichhardt}},
  \bibinfo {author} {\bibfnamefont {C.~D.}\ \bibnamefont {Batista}}, \ and\
  \bibinfo {author} {\bibfnamefont {A.}~\bibnamefont {Saxena}},\ }\href
  {\doibase 10.1103/PhysRevB.87.214419} {\bibfield  {journal} {\bibinfo
  {journal} {Phys. Rev. B}\ }\textbf {\bibinfo {volume} {87}},\ \bibinfo
  {pages} {214419} (\bibinfo {year} {2013})}\BibitemShut {NoStop}%
\bibitem [{\citenamefont {Iwasaki}\ \emph {et~al.}(2013)\citenamefont
  {Iwasaki}, \citenamefont {Mochizuki},\ and\ \citenamefont
  {Nagaosa}}]{Iwasaki2013}%
  \BibitemOpen
  \bibfield  {author} {\bibinfo {author} {\bibfnamefont {J.}~\bibnamefont
  {Iwasaki}}, \bibinfo {author} {\bibfnamefont {M.}~\bibnamefont {Mochizuki}},
  \ and\ \bibinfo {author} {\bibfnamefont {N.}~\bibnamefont {Nagaosa}},\ }\href
  {\doibase 10.1038/ncomms2442} {\bibfield  {journal} {\bibinfo  {journal}
  {Nat. Commun.}\ }\textbf {\bibinfo {volume} {4}},\ \bibinfo {pages} {1463}
  (\bibinfo {year} {2013})}\BibitemShut {NoStop}%
\bibitem [{\citenamefont {Lu}\ and\ \citenamefont {Xiang}(2014)}]{Lu2014}%
  \BibitemOpen
  \bibfield  {author} {\bibinfo {author} {\bibfnamefont {W.}~\bibnamefont
  {Lu}}\ and\ \bibinfo {author} {\bibfnamefont {J.}~\bibnamefont {Xiang}},\
  }\href@noop {} {\emph {\bibinfo {title} {Semiconductor Nanowires: From
  Next-Generation Electronics to Sustainable Energy}}}\ (\bibinfo  {publisher}
  {Royal Society of Chemistry},\ \bibinfo {year} {2014})\BibitemShut {NoStop}%
\bibitem [{\citenamefont {Koshibae}\ and\ \citenamefont
  {Nagaosa}(2016)}]{Koshibae2016}%
  \BibitemOpen
  \bibfield  {author} {\bibinfo {author} {\bibfnamefont {W.}~\bibnamefont
  {Koshibae}}\ and\ \bibinfo {author} {\bibfnamefont {N.}~\bibnamefont
  {Nagaosa}},\ }\href {\doibase 10.1038/ncomms10542} {\bibfield  {journal}
  {\bibinfo  {journal} {Nat. commun.}\ }\textbf {\bibinfo {volume} {7}},\
  \bibinfo {pages} {10542} (\bibinfo {year} {2016})}\BibitemShut {NoStop}%
\bibitem [{\citenamefont {Tchoe}\ and\ \citenamefont {Han}(2012)}]{Tchoe2012}%
  \BibitemOpen
  \bibfield  {author} {\bibinfo {author} {\bibfnamefont {Y.}~\bibnamefont
  {Tchoe}}\ and\ \bibinfo {author} {\bibfnamefont {J.~H.}\ \bibnamefont
  {Han}},\ }\href {\doibase 10.1103/PhysRevB.85.174416} {\bibfield  {journal}
  {\bibinfo  {journal} {Phys. Rev. B}\ }\textbf {\bibinfo {volume} {85}},\
  \bibinfo {pages} {174416} (\bibinfo {year} {2012})}\BibitemShut {NoStop}%
\bibitem [{\citenamefont {Romming}\ \emph {et~al.}(2013)\citenamefont
  {Romming}, \citenamefont {Hanneken}, \citenamefont {Menzel}, \citenamefont
  {Bickel}, \citenamefont {Wolter}, \citenamefont {von Bergmann}, \citenamefont
  {Kubetzka},\ and\ \citenamefont {Wiesendanger}}]{Romming2013}%
  \BibitemOpen
  \bibfield  {author} {\bibinfo {author} {\bibfnamefont {N.}~\bibnamefont
  {Romming}}, \bibinfo {author} {\bibfnamefont {C.}~\bibnamefont {Hanneken}},
  \bibinfo {author} {\bibfnamefont {M.}~\bibnamefont {Menzel}}, \bibinfo
  {author} {\bibfnamefont {J.~E.}\ \bibnamefont {Bickel}}, \bibinfo {author}
  {\bibfnamefont {B.}~\bibnamefont {Wolter}}, \bibinfo {author} {\bibfnamefont
  {K.}~\bibnamefont {von Bergmann}}, \bibinfo {author} {\bibfnamefont
  {A.}~\bibnamefont {Kubetzka}}, \ and\ \bibinfo {author} {\bibfnamefont
  {R.}~\bibnamefont {Wiesendanger}},\ }\href {\doibase 10.1126/science.1240573}
  {\bibfield  {journal} {\bibinfo  {journal} {Science}\ }\textbf {\bibinfo
  {volume} {341}},\ \bibinfo {pages} {636} (\bibinfo {year}
  {2013})}\BibitemShut {NoStop}%
\bibitem [{\citenamefont {Stier}\ \emph {et~al.}(2017)\citenamefont {Stier},
  \citenamefont {H\"ausler}, \citenamefont {Posske}, \citenamefont {Gurski},\
  and\ \citenamefont {Thorwart}}]{Stier2017}%
  \BibitemOpen
  \bibfield  {author} {\bibinfo {author} {\bibfnamefont {M.}~\bibnamefont
  {Stier}}, \bibinfo {author} {\bibfnamefont {W.}~\bibnamefont {H\"ausler}},
  \bibinfo {author} {\bibfnamefont {T.}~\bibnamefont {Posske}}, \bibinfo
  {author} {\bibfnamefont {G.}~\bibnamefont {Gurski}}, \ and\ \bibinfo {author}
  {\bibfnamefont {M.}~\bibnamefont {Thorwart}},\ }\href {\doibase
  10.1103/PhysRevLett.118.267203} {\bibfield  {journal} {\bibinfo  {journal}
  {Phys. Rev. Lett.}\ }\textbf {\bibinfo {volume} {118}},\ \bibinfo {pages}
  {267203} (\bibinfo {year} {2017})}\BibitemShut {NoStop}%
\bibitem [{\citenamefont {Everschor-Sitte}\ \emph {et~al.}(2017)\citenamefont
  {Everschor-Sitte}, \citenamefont {Sitte}, \citenamefont {Valet},
  \citenamefont {Abanov},\ and\ \citenamefont {Sinova}}]{Everschor_Sitte_2017}%
  \BibitemOpen
  \bibfield  {author} {\bibinfo {author} {\bibfnamefont {K.}~\bibnamefont
  {Everschor-Sitte}}, \bibinfo {author} {\bibfnamefont {M.}~\bibnamefont
  {Sitte}}, \bibinfo {author} {\bibfnamefont {T.}~\bibnamefont {Valet}},
  \bibinfo {author} {\bibfnamefont {A.}~\bibnamefont {Abanov}}, \ and\ \bibinfo
  {author} {\bibfnamefont {J.}~\bibnamefont {Sinova}},\ }\href@noop {}
  {\bibfield  {journal} {\bibinfo  {journal} {New J. Phys.}\ }\textbf {\bibinfo
  {volume} {19}},\ \bibinfo {pages} {092001} (\bibinfo {year}
  {2017})}\BibitemShut {NoStop}%
\bibitem [{\citenamefont {Müller}(2017)}]{Mueller_2017}%
  \BibitemOpen
  \bibfield  {author} {\bibinfo {author} {\bibfnamefont {J.}~\bibnamefont
  {Müller}},\ }\href@noop {} {\bibfield  {journal} {\bibinfo  {journal} {New
  J. Phys.}\ }\textbf {\bibinfo {volume} {19}},\ \bibinfo {pages} {025002}
  (\bibinfo {year} {2017})}\BibitemShut {NoStop}%
\bibitem [{\citenamefont {Suess}\ \emph {et~al.}(2018)\citenamefont {Suess},
  \citenamefont {Vogler}, \citenamefont {Bruckner}, \citenamefont
  {Heistracher},\ and\ \citenamefont {Abert}}]{Suess_2018}%
  \BibitemOpen
  \bibfield  {author} {\bibinfo {author} {\bibfnamefont {D.}~\bibnamefont
  {Suess}}, \bibinfo {author} {\bibfnamefont {C.}~\bibnamefont {Vogler}},
  \bibinfo {author} {\bibfnamefont {F.}~\bibnamefont {Bruckner}}, \bibinfo
  {author} {\bibfnamefont {P.}~\bibnamefont {Heistracher}}, \ and\ \bibinfo
  {author} {\bibfnamefont {C.}~\bibnamefont {Abert}},\ }\href@noop {}
  {\bibfield  {journal} {\bibinfo  {journal} {AIP Advances}\ }\textbf {\bibinfo
  {volume} {8}},\ \bibinfo {pages} {115301} (\bibinfo {year}
  {2018})}\BibitemShut {NoStop}%
\bibitem [{\citenamefont {Litzius}\ \emph {et~al.}(2017)\citenamefont
  {Litzius}, \citenamefont {Lemesh}, \citenamefont {Krüger}, \citenamefont
  {Bassirian}, \citenamefont {Caretta}, \citenamefont {Richter}, \citenamefont
  {Büttner}, \citenamefont {Sato}, \citenamefont {Tretiakov}, \citenamefont
  {Förster}, \citenamefont {Reeve}, \citenamefont {Weigand}, \citenamefont
  {Bykova}, \citenamefont {Stoll}, \citenamefont {Schütz}, \citenamefont
  {Beach},\ and\ \citenamefont {Kläui}}]{Litzius2017}%
  \BibitemOpen
  \bibfield  {author} {\bibinfo {author} {\bibfnamefont {K.}~\bibnamefont
  {Litzius}}, \bibinfo {author} {\bibfnamefont {I.}~\bibnamefont {Lemesh}},
  \bibinfo {author} {\bibfnamefont {B.}~\bibnamefont {Krüger}}, \bibinfo
  {author} {\bibfnamefont {P.}~\bibnamefont {Bassirian}}, \bibinfo {author}
  {\bibfnamefont {L.}~\bibnamefont {Caretta}}, \bibinfo {author} {\bibfnamefont
  {K.}~\bibnamefont {Richter}}, \bibinfo {author} {\bibfnamefont
  {F.}~\bibnamefont {Büttner}}, \bibinfo {author} {\bibfnamefont
  {K.}~\bibnamefont {Sato}}, \bibinfo {author} {\bibfnamefont {O.~A.}\
  \bibnamefont {Tretiakov}}, \bibinfo {author} {\bibfnamefont {J.}~\bibnamefont
  {Förster}}, \bibinfo {author} {\bibfnamefont {R.~M.}\ \bibnamefont {Reeve}},
  \bibinfo {author} {\bibfnamefont {M.}~\bibnamefont {Weigand}}, \bibinfo
  {author} {\bibfnamefont {I.}~\bibnamefont {Bykova}}, \bibinfo {author}
  {\bibfnamefont {H.}~\bibnamefont {Stoll}}, \bibinfo {author} {\bibfnamefont
  {G.}~\bibnamefont {Schütz}}, \bibinfo {author} {\bibfnamefont {G.~S.~D.}\
  \bibnamefont {Beach}}, \ and\ \bibinfo {author} {\bibfnamefont
  {M.}~\bibnamefont {Kläui}},\ }\href {\doibase 10.1038/nphys4000} {\bibfield
  {journal} {\bibinfo  {journal} {Nat. Phys.}\ }\textbf {\bibinfo {volume}
  {13}},\ \bibinfo {pages} {170} (\bibinfo {year} {2017})}\BibitemShut
  {NoStop}%
\bibitem [{\citenamefont {Jiang}\ \emph {et~al.}(2017)\citenamefont {Jiang},
  \citenamefont {Zhang}, \citenamefont {Yu}, \citenamefont {Zhang},
  \citenamefont {Wang}, \citenamefont {Jungfleisch}, \citenamefont {Pearson},
  \citenamefont {Cheng}, \citenamefont {Heinonen}, \citenamefont {Wang} \emph
  {et~al.}}]{Jiang2017}%
  \BibitemOpen
  \bibfield  {author} {\bibinfo {author} {\bibfnamefont {W.}~\bibnamefont
  {Jiang}}, \bibinfo {author} {\bibfnamefont {X.}~\bibnamefont {Zhang}},
  \bibinfo {author} {\bibfnamefont {G.}~\bibnamefont {Yu}}, \bibinfo {author}
  {\bibfnamefont {W.}~\bibnamefont {Zhang}}, \bibinfo {author} {\bibfnamefont
  {X.}~\bibnamefont {Wang}}, \bibinfo {author} {\bibfnamefont {M.~B.}\
  \bibnamefont {Jungfleisch}}, \bibinfo {author} {\bibfnamefont {J.~E.}\
  \bibnamefont {Pearson}}, \bibinfo {author} {\bibfnamefont {X.}~\bibnamefont
  {Cheng}}, \bibinfo {author} {\bibfnamefont {O.}~\bibnamefont {Heinonen}},
  \bibinfo {author} {\bibfnamefont {K.~L.}\ \bibnamefont {Wang}},  \emph
  {et~al.},\ }\href {\doibase http://dx.doi.org/10.1038/nphys3883} {\bibfield
  {journal} {\bibinfo  {journal} {Nat. Phys.}\ }\textbf {\bibinfo {volume}
  {13}},\ \bibinfo {pages} {162} (\bibinfo {year} {2017})}\BibitemShut
  {NoStop}%
\bibitem [{\citenamefont {Kolesnikov}\ \emph {et~al.}(2018)\citenamefont
  {Kolesnikov}, \citenamefont {Stebliy}, \citenamefont {Samardak},\ and\
  \citenamefont {Ognev}}]{Kolesnikov2018}%
  \BibitemOpen
  \bibfield  {author} {\bibinfo {author} {\bibfnamefont {A.~G.}\ \bibnamefont
  {Kolesnikov}}, \bibinfo {author} {\bibfnamefont {M.~E.}\ \bibnamefont
  {Stebliy}}, \bibinfo {author} {\bibfnamefont {A.~S.}\ \bibnamefont
  {Samardak}}, \ and\ \bibinfo {author} {\bibfnamefont {A.~V.}\ \bibnamefont
  {Ognev}},\ }\href {\doibase 10.1038/s41598-018-34934-2} {\bibfield  {journal}
  {\bibinfo  {journal} {Sci. Rep.}\ }\textbf {\bibinfo {volume} {8}},\ \bibinfo
  {pages} {16966} (\bibinfo {year} {2018})}\BibitemShut {NoStop}%
\bibitem [{\citenamefont {Lew Dawidowitsch~Landau}(1935)}]{Landau1935}%
  \BibitemOpen
  \bibfield  {author} {\bibinfo {author} {\bibfnamefont {J.~M.~L.}\
  \bibnamefont {Lew Dawidowitsch~Landau}},\ }\href {\doibase
  10.1016/B978-0-08-010586-4.50023-7} {\bibfield  {journal} {\bibinfo
  {journal} {Phys. Z. Sowj.}\ ,\ \bibinfo {pages} {153}} (\bibinfo {year}
  {1935})}\BibitemShut {NoStop}%
\bibitem [{\citenamefont {Landau}\ and\ \citenamefont
  {Lifshitz}(1960)}]{Landau1960}%
  \BibitemOpen
  \bibfield  {author} {\bibinfo {author} {\bibfnamefont {L.~D.}\ \bibnamefont
  {Landau}}\ and\ \bibinfo {author} {\bibfnamefont {E.~M.}\ \bibnamefont
  {Lifshitz}},\ }\href {\doibase 10.1143/JJAP.28.2485} {\emph {\bibinfo {title}
  {Electrodynamics of continuous media}}}\ (\bibinfo  {publisher} {Pergamon
  Press},\ \bibinfo {year} {1960})\BibitemShut {NoStop}%
\bibitem [{\citenamefont {Gilbert}(2004)}]{Gilbert2004}%
  \BibitemOpen
  \bibfield  {author} {\bibinfo {author} {\bibfnamefont {T.~L.}\ \bibnamefont
  {Gilbert}},\ }\href {\doibase 10.1109/TMAG.2004.836740} {\bibfield  {journal}
  {\bibinfo  {journal} {IEEE Trans. Magn.}\ }\textbf {\bibinfo {volume} {40}},\
  \bibinfo {pages} {3443} (\bibinfo {year} {2004})}\BibitemShut {NoStop}%
\bibitem [{\citenamefont {Nakatani}\ \emph {et~al.}(1989)\citenamefont
  {Nakatani}, \citenamefont {Uesaka},\ and\ \citenamefont
  {Hayashi}}]{Nakatani1989}%
  \BibitemOpen
  \bibfield  {author} {\bibinfo {author} {\bibfnamefont {Y.}~\bibnamefont
  {Nakatani}}, \bibinfo {author} {\bibfnamefont {Y.}~\bibnamefont {Uesaka}}, \
  and\ \bibinfo {author} {\bibfnamefont {N.}~\bibnamefont {Hayashi}},\ }\href
  {\doibase 10.1143/JJAP.28.2485} {\bibfield  {journal} {\bibinfo  {journal}
  {Jpn. J. Appl. Phys}\ }\textbf {\bibinfo {volume} {28}},\ \bibinfo {pages}
  {2485} (\bibinfo {year} {1989})}\BibitemShut {NoStop}%
\bibitem [{\citenamefont {Slonczewski}(1996)}]{Slonczewski1996}%
  \BibitemOpen
  \bibfield  {author} {\bibinfo {author} {\bibfnamefont {J.~C.}\ \bibnamefont
  {Slonczewski}},\ }\href {\doibase 10.1016/0304-8853(96)00062-5} {\bibfield
  {journal} {\bibinfo  {journal} {J. Magn. Magn. Mater.}\ }\textbf {\bibinfo
  {volume} {159}},\ \bibinfo {pages} {L1} (\bibinfo {year} {1996})}\BibitemShut
  {NoStop}%
\bibitem [{\citenamefont {Bazaliy}\ \emph {et~al.}(1998)\citenamefont
  {Bazaliy}, \citenamefont {Jones},\ and\ \citenamefont {Zhang}}]{Bazaliy1998}%
  \BibitemOpen
  \bibfield  {author} {\bibinfo {author} {\bibfnamefont {Y.~B.}\ \bibnamefont
  {Bazaliy}}, \bibinfo {author} {\bibfnamefont {B.~A.}\ \bibnamefont {Jones}},
  \ and\ \bibinfo {author} {\bibfnamefont {S.-C.}\ \bibnamefont {Zhang}},\
  }\href {\doibase 10.1103/PhysRevB.57.R3213} {\bibfield  {journal} {\bibinfo
  {journal} {Phys. Rev. B}\ }\textbf {\bibinfo {volume} {57}},\ \bibinfo
  {pages} {R3213} (\bibinfo {year} {1998})}\BibitemShut {NoStop}%
\bibitem [{\citenamefont {Zhang}\ and\ \citenamefont {Li}(2004)}]{Zhang2004}%
  \BibitemOpen
  \bibfield  {author} {\bibinfo {author} {\bibfnamefont {S.}~\bibnamefont
  {Zhang}}\ and\ \bibinfo {author} {\bibfnamefont {Z.}~\bibnamefont {Li}},\
  }\href {\doibase 10.1103/PhysRevLett.93.127204} {\bibfield  {journal}
  {\bibinfo  {journal} {Phys. Rev. Lett.}\ }\textbf {\bibinfo {volume} {93}},\
  \bibinfo {pages} {127204} (\bibinfo {year} {2004})}\BibitemShut {NoStop}%
\bibitem [{SM()}]{SM}%
  \BibitemOpen
  \href@noop {} {}\bibinfo {note} {See Supplemental Material in the ancillary folder of arXiv for the calculation of first and second order spin torques.  Furthermore, details about the Laplace equation and videos of the skyrmion steering are provided.  The Supplemental Material includes Refs. \cite{bychkov1984,Borge2014}}\BibitemShut
  {NoStop}%
\bibitem [{\citenamefont {Manchon}\ and\ \citenamefont
  {Zhang}(2009)}]{Manchon2009}%
  \BibitemOpen
  \bibfield  {author} {\bibinfo {author} {\bibfnamefont {A.}~\bibnamefont
  {Manchon}}\ and\ \bibinfo {author} {\bibfnamefont {S.}~\bibnamefont
  {Zhang}},\ }\href {\doibase 10.1103/PhysRevB.79.094422} {\bibfield  {journal}
  {\bibinfo  {journal} {Phys. Rev. B}\ }\textbf {\bibinfo {volume} {79}},\
  \bibinfo {pages} {094422} (\bibinfo {year} {2009})}\BibitemShut {NoStop}%
\bibitem [{\citenamefont {van~der Bijl}\ and\ \citenamefont
  {Duine}(2012)}]{vanderBijl2012}%
  \BibitemOpen
  \bibfield  {author} {\bibinfo {author} {\bibfnamefont {E.}~\bibnamefont
  {van~der Bijl}}\ and\ \bibinfo {author} {\bibfnamefont {R.~A.}\ \bibnamefont
  {Duine}},\ }\href {\doibase 10.1103/PhysRevB.86.094406} {\bibfield  {journal}
  {\bibinfo  {journal} {Phys. Rev. B}\ }\textbf {\bibinfo {volume} {86}},\
  \bibinfo {pages} {094406} (\bibinfo {year} {2012})}\BibitemShut {NoStop}%
\bibitem [{\citenamefont {Stier}\ \emph {et~al.}(2013)\citenamefont {Stier},
  \citenamefont {Egger},\ and\ \citenamefont {Thorwart}}]{Stier2013}%
  \BibitemOpen
  \bibfield  {author} {\bibinfo {author} {\bibfnamefont {M.}~\bibnamefont
  {Stier}}, \bibinfo {author} {\bibfnamefont {R.}~\bibnamefont {Egger}}, \ and\
  \bibinfo {author} {\bibfnamefont {M.}~\bibnamefont {Thorwart}},\ }\href
  {\doibase 10.1103/PhysRevB.87.184415} {\bibfield  {journal} {\bibinfo
  {journal} {Phys. Rev. B}\ }\textbf {\bibinfo {volume} {87}},\ \bibinfo
  {pages} {184415} (\bibinfo {year} {2013})}\BibitemShut {NoStop}%
\bibitem [{\citenamefont {Stier}\ \emph {et~al.}(2014)\citenamefont {Stier},
  \citenamefont {Creutzburg},\ and\ \citenamefont {Thorwart}}]{Stier2014}%
  \BibitemOpen
  \bibfield  {author} {\bibinfo {author} {\bibfnamefont {M.}~\bibnamefont
  {Stier}}, \bibinfo {author} {\bibfnamefont {M.}~\bibnamefont {Creutzburg}}, \
  and\ \bibinfo {author} {\bibfnamefont {M.}~\bibnamefont {Thorwart}},\ }\href
  {\doibase 10.1103/PhysRevB.90.014433} {\bibfield  {journal} {\bibinfo
  {journal} {Phys. Rev. B}\ }\textbf {\bibinfo {volume} {90}},\ \bibinfo
  {pages} {014433} (\bibinfo {year} {2014})}\BibitemShut {NoStop}%
\bibitem [{\citenamefont {Stier}\ and\ \citenamefont
  {Thorwart}(2015)}]{Stier2015}%
  \BibitemOpen
  \bibfield  {author} {\bibinfo {author} {\bibfnamefont {M.}~\bibnamefont
  {Stier}}\ and\ \bibinfo {author} {\bibfnamefont {M.}~\bibnamefont
  {Thorwart}},\ }\href {\doibase 10.1103/PhysRevB.92.220406} {\bibfield
  {journal} {\bibinfo  {journal} {Phys. Rev. B}\ }\textbf {\bibinfo {volume}
  {92}},\ \bibinfo {pages} {220406} (\bibinfo {year} {2015})}\BibitemShut
  {NoStop}%
\bibitem [{\citenamefont {Thomas}\ \emph {et~al.}(2006)\citenamefont {Thomas},
  \citenamefont {Hayashi}, \citenamefont {Jiang}, \citenamefont {Moriya},
  \citenamefont {Rettner},\ and\ \citenamefont {Parkin}}]{Thomas2006}%
  \BibitemOpen
  \bibfield  {author} {\bibinfo {author} {\bibfnamefont {L.}~\bibnamefont
  {Thomas}}, \bibinfo {author} {\bibfnamefont {M.}~\bibnamefont {Hayashi}},
  \bibinfo {author} {\bibfnamefont {X.}~\bibnamefont {Jiang}}, \bibinfo
  {author} {\bibfnamefont {R.}~\bibnamefont {Moriya}}, \bibinfo {author}
  {\bibfnamefont {C.}~\bibnamefont {Rettner}}, \ and\ \bibinfo {author}
  {\bibfnamefont {S.~S.}\ \bibnamefont {Parkin}},\ }\href {\doibase
  10.1038/nature05093} {\bibfield  {journal} {\bibinfo  {journal} {Nature}\
  }\textbf {\bibinfo {volume} {443}},\ \bibinfo {pages} {197} (\bibinfo {year}
  {2006})}\BibitemShut {NoStop}%
\bibitem [{\citenamefont {K{\"o}tzler}\ \emph {et~al.}(2007)\citenamefont
  {K{\"o}tzler}, \citenamefont {G{\"o}rlitz},\ and\ \citenamefont
  {Wiekhorst}}]{Koetzler2007}%
  \BibitemOpen
  \bibfield  {author} {\bibinfo {author} {\bibfnamefont {J.}~\bibnamefont
  {K{\"o}tzler}}, \bibinfo {author} {\bibfnamefont {D.}~\bibnamefont
  {G{\"o}rlitz}}, \ and\ \bibinfo {author} {\bibfnamefont {F.}~\bibnamefont
  {Wiekhorst}},\ }\href {\doibase 10.1103/PhysRevB.76.104404} {\bibfield
  {journal} {\bibinfo  {journal} {Phys. Rev. B}\ }\textbf {\bibinfo {volume}
  {76}},\ \bibinfo {pages} {104404} (\bibinfo {year} {2007})}\BibitemShut
  {NoStop}%
\bibitem [{\citenamefont {Sekiguchi}\ \emph {et~al.}(2012)\citenamefont
  {Sekiguchi}, \citenamefont {Yamada}, \citenamefont {Seo}, \citenamefont
  {Lee}, \citenamefont {Chiba}, \citenamefont {Kobayashi},\ and\ \citenamefont
  {Ono}}]{Sekiguchi2012}%
  \BibitemOpen
  \bibfield  {author} {\bibinfo {author} {\bibfnamefont {K.}~\bibnamefont
  {Sekiguchi}}, \bibinfo {author} {\bibfnamefont {K.}~\bibnamefont {Yamada}},
  \bibinfo {author} {\bibfnamefont {S.-M.}\ \bibnamefont {Seo}}, \bibinfo
  {author} {\bibfnamefont {K.-J.}\ \bibnamefont {Lee}}, \bibinfo {author}
  {\bibfnamefont {D.}~\bibnamefont {Chiba}}, \bibinfo {author} {\bibfnamefont
  {K.}~\bibnamefont {Kobayashi}}, \ and\ \bibinfo {author} {\bibfnamefont
  {T.}~\bibnamefont {Ono}},\ }\href {\doibase 10.1103/PhysRevLett.108.017203}
  {\bibfield  {journal} {\bibinfo  {journal} {Phys. Rev. Lett.}\ }\textbf
  {\bibinfo {volume} {108}},\ \bibinfo {pages} {017203} (\bibinfo {year}
  {2012})}\BibitemShut {NoStop}%
\bibitem [{\citenamefont {Martinez}\ \emph {et~al.}(2008)\citenamefont
  {Martinez}, \citenamefont {Lopez-Diaz}, \citenamefont {Alejos},\ and\
  \citenamefont {Torres}}]{Martinez2008}%
  \BibitemOpen
  \bibfield  {author} {\bibinfo {author} {\bibfnamefont {E.}~\bibnamefont
  {Martinez}}, \bibinfo {author} {\bibfnamefont {L.}~\bibnamefont
  {Lopez-Diaz}}, \bibinfo {author} {\bibfnamefont {O.}~\bibnamefont {Alejos}},
  \ and\ \bibinfo {author} {\bibfnamefont {L.}~\bibnamefont {Torres}},\ }\href
  {\doibase 10.1103/PhysRevB.77.144417} {\bibfield  {journal} {\bibinfo
  {journal} {Phys. Rev. B}\ }\textbf {\bibinfo {volume} {77}},\ \bibinfo
  {pages} {144417} (\bibinfo {year} {2008})}\BibitemShut {NoStop}%
\bibitem [{\citenamefont {Ho~Park}\ \emph {et~al.}(2013)\citenamefont
  {Ho~Park}, \citenamefont {Kim}, \citenamefont {Chang}, \citenamefont
  {Hee~Han}, \citenamefont {Eom}, \citenamefont {Choi},\ and\ \citenamefont
  {Cheol~Koo}}]{Ho2013}%
  \BibitemOpen
  \bibfield  {author} {\bibinfo {author} {\bibfnamefont {Y.}~\bibnamefont
  {Ho~Park}}, \bibinfo {author} {\bibfnamefont {H.-j.}\ \bibnamefont {Kim}},
  \bibinfo {author} {\bibfnamefont {J.}~\bibnamefont {Chang}}, \bibinfo
  {author} {\bibfnamefont {S.}~\bibnamefont {Hee~Han}}, \bibinfo {author}
  {\bibfnamefont {J.}~\bibnamefont {Eom}}, \bibinfo {author} {\bibfnamefont
  {H.-J.}\ \bibnamefont {Choi}}, \ and\ \bibinfo {author} {\bibfnamefont
  {H.}~\bibnamefont {Cheol~Koo}},\ }\href {\doibase 10.1063/1.4855495}
  {\bibfield  {journal} {\bibinfo  {journal} {Appl. Phys. Lett.}\ }\textbf
  {\bibinfo {volume} {103}},\ \bibinfo {pages} {252407} (\bibinfo {year}
  {2013})}\BibitemShut {NoStop}%
\bibitem [{\citenamefont {Caviglia}\ \emph {et~al.}(2010)\citenamefont
  {Caviglia}, \citenamefont {Gabay}, \citenamefont {Gariglio}, \citenamefont
  {Reyren}, \citenamefont {Cancellieri},\ and\ \citenamefont
  {Triscone}}]{Caviglia2010_tunableRashba}%
  \BibitemOpen
  \bibfield  {author} {\bibinfo {author} {\bibfnamefont {A.~D.}\ \bibnamefont
  {Caviglia}}, \bibinfo {author} {\bibfnamefont {M.}~\bibnamefont {Gabay}},
  \bibinfo {author} {\bibfnamefont {S.}~\bibnamefont {Gariglio}}, \bibinfo
  {author} {\bibfnamefont {N.}~\bibnamefont {Reyren}}, \bibinfo {author}
  {\bibfnamefont {C.}~\bibnamefont {Cancellieri}}, \ and\ \bibinfo {author}
  {\bibfnamefont {J.-M.}\ \bibnamefont {Triscone}},\ }\href {\doibase
  10.1103/PhysRevLett.104.126803} {\bibfield  {journal} {\bibinfo  {journal}
  {Phys. Rev. Lett.}\ }\textbf {\bibinfo {volume} {104}},\ \bibinfo {pages}
  {126803} (\bibinfo {year} {2010})}\BibitemShut {NoStop}%
\bibitem [{\citenamefont {Chen}\ \emph {et~al.}(2018)\citenamefont {Chen},
  \citenamefont {Gmitra}, \citenamefont {Vogel}, \citenamefont {Islinger},
  \citenamefont {Kronseder}, \citenamefont {Schuh}, \citenamefont {Bougeard},
  \citenamefont {Fabian}, \citenamefont {Weiss},\ and\ \citenamefont
  {Back}}]{chen2018electric}%
  \BibitemOpen
  \bibfield  {author} {\bibinfo {author} {\bibfnamefont {L.}~\bibnamefont
  {Chen}}, \bibinfo {author} {\bibfnamefont {M.}~\bibnamefont {Gmitra}},
  \bibinfo {author} {\bibfnamefont {M.}~\bibnamefont {Vogel}}, \bibinfo
  {author} {\bibfnamefont {R.}~\bibnamefont {Islinger}}, \bibinfo {author}
  {\bibfnamefont {M.}~\bibnamefont {Kronseder}}, \bibinfo {author}
  {\bibfnamefont {D.}~\bibnamefont {Schuh}}, \bibinfo {author} {\bibfnamefont
  {D.}~\bibnamefont {Bougeard}}, \bibinfo {author} {\bibfnamefont
  {J.}~\bibnamefont {Fabian}}, \bibinfo {author} {\bibfnamefont
  {D.}~\bibnamefont {Weiss}}, \ and\ \bibinfo {author} {\bibfnamefont
  {C.}~\bibnamefont {Back}},\ }\href@noop {} {\bibfield  {journal} {\bibinfo
  {journal} {arXiv preprint arXiv:1803.01656}\ } (\bibinfo {year}
  {2018})}\BibitemShut {NoStop}%
\bibitem [{\citenamefont {Emori}\ \emph {et~al.}(2014)\citenamefont {Emori},
  \citenamefont {Bauer}, \citenamefont {Woo},\ and\ \citenamefont
  {Beach}}]{soimodification2014}%
  \BibitemOpen
  \bibfield  {author} {\bibinfo {author} {\bibfnamefont {S.}~\bibnamefont
  {Emori}}, \bibinfo {author} {\bibfnamefont {U.}~\bibnamefont {Bauer}},
  \bibinfo {author} {\bibfnamefont {S.}~\bibnamefont {Woo}}, \ and\ \bibinfo
  {author} {\bibfnamefont {G.~S.~D.}\ \bibnamefont {Beach}},\ }\href@noop {}
  {\bibfield  {journal} {\bibinfo  {journal} {Applied Physics Letters}\
  }\textbf {\bibinfo {volume} {105}},\ \bibinfo {pages} {222401} (\bibinfo
  {year} {2014})}\BibitemShut {NoStop}%
\bibitem [{\citenamefont {Yin}\ \emph {et~al.}(2019)\citenamefont {Yin},
  \citenamefont {Seiler}, \citenamefont {Tang}, \citenamefont {Leermakers},
  \citenamefont {Lebedev}, \citenamefont {Zeitler},\ and\ \citenamefont
  {Aarts}}]{yin2019tuning}%
  \BibitemOpen
  \bibfield  {author} {\bibinfo {author} {\bibfnamefont {C.}~\bibnamefont
  {Yin}}, \bibinfo {author} {\bibfnamefont {P.}~\bibnamefont {Seiler}},
  \bibinfo {author} {\bibfnamefont {L.~M.}\ \bibnamefont {Tang}}, \bibinfo
  {author} {\bibfnamefont {I.}~\bibnamefont {Leermakers}}, \bibinfo {author}
  {\bibfnamefont {N.}~\bibnamefont {Lebedev}}, \bibinfo {author} {\bibfnamefont
  {U.}~\bibnamefont {Zeitler}}, \ and\ \bibinfo {author} {\bibfnamefont
  {J.}~\bibnamefont {Aarts}},\ }\href@noop {} {\bibfield  {journal} {\bibinfo
  {journal} {arXiv preprint arXiv:1904.03731}\ } (\bibinfo {year}
  {2019})}\BibitemShut {NoStop}%
\bibitem [{\citenamefont {Narayanapillai}\ \emph {et~al.}(2014)\citenamefont
  {Narayanapillai}, \citenamefont {Gopinadhan}, \citenamefont {Qiu},
  \citenamefont {Annadi}, \citenamefont {Ariando}, \citenamefont {Venkatesan},\
  and\ \citenamefont {Yang}}]{narayanapillai2014current}%
  \BibitemOpen
  \bibfield  {author} {\bibinfo {author} {\bibfnamefont {K.}~\bibnamefont
  {Narayanapillai}}, \bibinfo {author} {\bibfnamefont {K.}~\bibnamefont
  {Gopinadhan}}, \bibinfo {author} {\bibfnamefont {X.}~\bibnamefont {Qiu}},
  \bibinfo {author} {\bibfnamefont {A.}~\bibnamefont {Annadi}}, \bibinfo
  {author} {\bibnamefont {Ariando}}, \bibinfo {author} {\bibfnamefont
  {T.}~\bibnamefont {Venkatesan}}, \ and\ \bibinfo {author} {\bibfnamefont
  {H.}~\bibnamefont {Yang}},\ }\href@noop {} {\bibfield  {journal} {\bibinfo
  {journal} {Applied Physics Letters}\ }\textbf {\bibinfo {volume} {105}},\
  \bibinfo {pages} {162405} (\bibinfo {year} {2014})}\BibitemShut {NoStop}%
\bibitem [{\citenamefont {Lin}\ \emph {et~al.}(2019)\citenamefont {Lin},
  \citenamefont {Li}, \citenamefont {Do{\u{g}}an}, \citenamefont {Li},
  \citenamefont {Rotella}, \citenamefont {Yu}, \citenamefont {Zhang},
  \citenamefont {Li}, \citenamefont {Lew}, \citenamefont {Wang} \emph
  {et~al.}}]{lin2019interface}%
  \BibitemOpen
  \bibfield  {author} {\bibinfo {author} {\bibfnamefont {W.}~\bibnamefont
  {Lin}}, \bibinfo {author} {\bibfnamefont {L.}~\bibnamefont {Li}}, \bibinfo
  {author} {\bibfnamefont {F.}~\bibnamefont {Do{\u{g}}an}}, \bibinfo {author}
  {\bibfnamefont {C.}~\bibnamefont {Li}}, \bibinfo {author} {\bibfnamefont
  {H.}~\bibnamefont {Rotella}}, \bibinfo {author} {\bibfnamefont
  {X.}~\bibnamefont {Yu}}, \bibinfo {author} {\bibfnamefont {B.}~\bibnamefont
  {Zhang}}, \bibinfo {author} {\bibfnamefont {Y.}~\bibnamefont {Li}}, \bibinfo
  {author} {\bibfnamefont {W.~S.}\ \bibnamefont {Lew}}, \bibinfo {author}
  {\bibfnamefont {S.}~\bibnamefont {Wang}},  \emph {et~al.},\ }\href@noop {}
  {\bibfield  {journal} {\bibinfo  {journal} {Nature communications}\ }\textbf
  {\bibinfo {volume} {10}},\ \bibinfo {pages} {1} (\bibinfo {year}
  {2019})}\BibitemShut {NoStop}%
\bibitem [{\citenamefont {Lesne}\ \emph {et~al.}(2016)\citenamefont {Lesne},
  \citenamefont {Fu}, \citenamefont {Oyarzun}, \citenamefont
  {Rojas-S{\'a}nchez}, \citenamefont {Vaz}, \citenamefont {Naganuma},
  \citenamefont {Sicoli}, \citenamefont {Attan{\'e}}, \citenamefont {Jamet},
  \citenamefont {Jacquet} \emph {et~al.}}]{lesne2016highly}%
  \BibitemOpen
  \bibfield  {author} {\bibinfo {author} {\bibfnamefont {E.}~\bibnamefont
  {Lesne}}, \bibinfo {author} {\bibfnamefont {Y.}~\bibnamefont {Fu}}, \bibinfo
  {author} {\bibfnamefont {S.}~\bibnamefont {Oyarzun}}, \bibinfo {author}
  {\bibfnamefont {J.}~\bibnamefont {Rojas-S{\'a}nchez}}, \bibinfo {author}
  {\bibfnamefont {D.}~\bibnamefont {Vaz}}, \bibinfo {author} {\bibfnamefont
  {H.}~\bibnamefont {Naganuma}}, \bibinfo {author} {\bibfnamefont
  {G.}~\bibnamefont {Sicoli}}, \bibinfo {author} {\bibfnamefont {J.-P.}\
  \bibnamefont {Attan{\'e}}}, \bibinfo {author} {\bibfnamefont
  {M.}~\bibnamefont {Jamet}}, \bibinfo {author} {\bibfnamefont
  {E.}~\bibnamefont {Jacquet}},  \emph {et~al.},\ }\href@noop {} {\bibfield
  {journal} {\bibinfo  {journal} {Nature materials}\ }\textbf {\bibinfo
  {volume} {15}},\ \bibinfo {pages} {1261} (\bibinfo {year}
  {2016})}\BibitemShut {NoStop}%
\bibitem [{\citenamefont {Langner}(2009)}]{langner2009ultrafast}%
  \BibitemOpen
  \bibfield  {author} {\bibinfo {author} {\bibfnamefont {M.~C.}\ \bibnamefont
  {Langner}},\ }\emph {\bibinfo {title} {Ultrafast Magnetization Dynamics of
  SrRuO 3 Thin Films}},\ \href@noop {} {Ph.D. thesis},\ \bibinfo  {school}
  {University of California, Berkeley} (\bibinfo {year} {2009})\BibitemShut
  {NoStop}%
\bibitem [{\citenamefont {Koster}\ \emph {et~al.}(2012)\citenamefont {Koster},
  \citenamefont {Klein}, \citenamefont {Siemons}, \citenamefont {Rijnders},
  \citenamefont {Dodge}, \citenamefont {Eom}, \citenamefont {Blank},\ and\
  \citenamefont {Beasley}}]{RevModPhys.84.253}%
  \BibitemOpen
  \bibfield  {author} {\bibinfo {author} {\bibfnamefont {G.}~\bibnamefont
  {Koster}}, \bibinfo {author} {\bibfnamefont {L.}~\bibnamefont {Klein}},
  \bibinfo {author} {\bibfnamefont {W.}~\bibnamefont {Siemons}}, \bibinfo
  {author} {\bibfnamefont {G.}~\bibnamefont {Rijnders}}, \bibinfo {author}
  {\bibfnamefont {J.~S.}\ \bibnamefont {Dodge}}, \bibinfo {author}
  {\bibfnamefont {C.-B.}\ \bibnamefont {Eom}}, \bibinfo {author} {\bibfnamefont
  {D.~H.~A.}\ \bibnamefont {Blank}}, \ and\ \bibinfo {author} {\bibfnamefont
  {M.~R.}\ \bibnamefont {Beasley}},\ }\href {\doibase
  10.1103/RevModPhys.84.253} {\bibfield  {journal} {\bibinfo  {journal} {Rev.
  Mod. Phys.}\ }\textbf {\bibinfo {volume} {84}},\ \bibinfo {pages} {253}
  (\bibinfo {year} {2012})}\BibitemShut {NoStop}%
\bibitem [{\citenamefont {Meng}\ \emph {et~al.}(2019)\citenamefont {Meng},
  \citenamefont {Ahmed}, \citenamefont {Baćani}, \citenamefont {Mandru},
  \citenamefont {Zhao}, \citenamefont {Bagués}, \citenamefont {Esser},
  \citenamefont {Flores}, \citenamefont {McComb}, \citenamefont {Hug},\ and\
  \citenamefont {Yang}}]{nanolett19}%
  \BibitemOpen
  \bibfield  {author} {\bibinfo {author} {\bibfnamefont {K.-Y.}\ \bibnamefont
  {Meng}}, \bibinfo {author} {\bibfnamefont {A.~S.}\ \bibnamefont {Ahmed}},
  \bibinfo {author} {\bibfnamefont {M.}~\bibnamefont {Baćani}}, \bibinfo
  {author} {\bibfnamefont {A.-O.}\ \bibnamefont {Mandru}}, \bibinfo {author}
  {\bibfnamefont {X.}~\bibnamefont {Zhao}}, \bibinfo {author} {\bibfnamefont
  {N.}~\bibnamefont {Bagués}}, \bibinfo {author} {\bibfnamefont {B.~D.}\
  \bibnamefont {Esser}}, \bibinfo {author} {\bibfnamefont {J.}~\bibnamefont
  {Flores}}, \bibinfo {author} {\bibfnamefont {D.~W.}\ \bibnamefont {McComb}},
  \bibinfo {author} {\bibfnamefont {H.~J.}\ \bibnamefont {Hug}}, \ and\
  \bibinfo {author} {\bibfnamefont {F.}~\bibnamefont {Yang}},\ }\href {\doibase
  10.1021/acs.nanolett.9b00596} {\bibfield  {journal} {\bibinfo  {journal}
  {Nano Letters}\ }\textbf {\bibinfo {volume} {19}},\ \bibinfo {pages} {3169}
  (\bibinfo {year} {2019})},\ \bibinfo {note} {pMID: 30935207},\ \Eprint
  {http://arxiv.org/abs/https://doi.org/10.1021/acs.nanolett.9b00596}
  {https://doi.org/10.1021/acs.nanolett.9b00596} \BibitemShut {NoStop}%
\bibitem [{\citenamefont {Ohuchi}\ \emph {et~al.}(2018)\citenamefont {Ohuchi},
  \citenamefont {Matsuno}, \citenamefont {Ogawa}, \citenamefont {Kozuka},
  \citenamefont {Uchida}, \citenamefont {Tokura},\ and\ \citenamefont
  {Kawasaki}}]{ohuchi2018electric}%
  \BibitemOpen
  \bibfield  {author} {\bibinfo {author} {\bibfnamefont {Y.}~\bibnamefont
  {Ohuchi}}, \bibinfo {author} {\bibfnamefont {J.}~\bibnamefont {Matsuno}},
  \bibinfo {author} {\bibfnamefont {N.}~\bibnamefont {Ogawa}}, \bibinfo
  {author} {\bibfnamefont {Y.}~\bibnamefont {Kozuka}}, \bibinfo {author}
  {\bibfnamefont {M.}~\bibnamefont {Uchida}}, \bibinfo {author} {\bibfnamefont
  {Y.}~\bibnamefont {Tokura}}, \ and\ \bibinfo {author} {\bibfnamefont
  {M.}~\bibnamefont {Kawasaki}},\ }\href@noop {} {\bibfield  {journal}
  {\bibinfo  {journal} {Nature communications}\ }\textbf {\bibinfo {volume}
  {9}},\ \bibinfo {pages} {1} (\bibinfo {year} {2018})}\BibitemShut {NoStop}%
\bibitem [{\citenamefont {Gu}\ \emph {et~al.}(2018)\citenamefont {Gu},
  \citenamefont {Wei}, \citenamefont {Xu}, \citenamefont {Zhang}, \citenamefont
  {Wang}, \citenamefont {Li}, \citenamefont {Saleem}, \citenamefont {Chang},
  \citenamefont {Sun}, \citenamefont {Song} \emph {et~al.}}]{gu2018oxygen}%
  \BibitemOpen
  \bibfield  {author} {\bibinfo {author} {\bibfnamefont {Y.}~\bibnamefont
  {Gu}}, \bibinfo {author} {\bibfnamefont {Y.-W.}\ \bibnamefont {Wei}},
  \bibinfo {author} {\bibfnamefont {K.}~\bibnamefont {Xu}}, \bibinfo {author}
  {\bibfnamefont {H.}~\bibnamefont {Zhang}}, \bibinfo {author} {\bibfnamefont
  {F.}~\bibnamefont {Wang}}, \bibinfo {author} {\bibfnamefont {F.}~\bibnamefont
  {Li}}, \bibinfo {author} {\bibfnamefont {M.~S.}\ \bibnamefont {Saleem}},
  \bibinfo {author} {\bibfnamefont {C.-Z.}\ \bibnamefont {Chang}}, \bibinfo
  {author} {\bibfnamefont {J.}~\bibnamefont {Sun}}, \bibinfo {author}
  {\bibfnamefont {C.}~\bibnamefont {Song}},  \emph {et~al.},\ }\href@noop {}
  {\bibfield  {journal} {\bibinfo  {journal} {arXiv preprint arXiv:1811.09075}\
  } (\bibinfo {year} {2018})}\BibitemShut {NoStop}%
\bibitem [{\citenamefont {Bychkov}\ and\ \citenamefont
  {Rashba}(1984)}]{bychkov1984}%
  \BibitemOpen
  \bibfield  {author} {\bibinfo {author} {\bibfnamefont { Y.~A. }\ \bibnamefont
  {Bychkov}}\ and\ \bibinfo {author} {\bibfnamefont {E.~I.}\ \bibnamefont
  {Rashba}},\ }{\bibfield
  {journal} {\bibinfo  {journal} {J. Phys. C}\ }\textbf {\bibinfo
  {volume} {17}},\ \bibinfo {pages} {6039} (\bibinfo {year} {1984})}\BibitemShut
  {NoStop}%
\bibitem [{\citenamefont {Borge}\ \emph {et~al.}(2014)\citenamefont {Borge},
  \citenamefont {Gorini}, \citenamefont {Vignale},\ and\ \citenamefont
  {Raimondi}}]{Borge2014}%
  \BibitemOpen
  \bibfield  {author} {\bibinfo {author} {\bibfnamefont {J.}~\bibnamefont
  {Borge}}, \bibinfo {author} {\bibfnamefont {C.}~\bibnamefont {Gorini}},
  \bibinfo {author} {\bibfnamefont {G.}~\bibnamefont {Vignale}}, \ and\
  \bibinfo {author} {\bibfnamefont {R.}~\bibnamefont {Raimondi}},\ }\href
  {\doibase 10.1103/PhysRevB.89.245443} {\bibfield  {journal} {\bibinfo
  {journal} {Phys. Rev. B}\ }\textbf {\bibinfo {volume} {89}} (\bibinfo {year}
  {2014}),\ 10.1103/PhysRevB.89.245443}\BibitemShut {NoStop}%
\end{thebibliography}

%

\end{document}